\documentclass[preprint,showpacs,preprintnumbers,amsmath,amssymb]{revtex4}
\usepackage[utf8]{inputenc}

\usepackage{graphicx}
\usepackage{psfrag}
\usepackage{amsmath}

\begin{document}

\preprint{TUD-ITP-TQO/04-2010-V100303}

\title{Analytical results for Josephson dynamics of ultracold Bosons}

\author{Lena Simon}
     \affiliation{Institut f\"{u}r Theoretische Physik,
Technische Universit\"at Dresden, D-01062 Dresden, Germany}
\author{Walter T. Strunz}
   \affiliation{Institut f\"{u}r Theoretische Physik,
Technische Universit\"at Dresden, D-01062 Dresden, Germany}

\date{\today}

\begin{abstract}
We study the dynamics of ultracold Bosons in a double-well potential within the 
two-mode Bose-Hubbard model by means of semiclassical methods. 
By applying a WKB quantization we find analytical
results for the energy spectrum, which are in excellent agreement 
with numerical exact results. They are valid in the energy range of plasma oscillations,
both in the Rabi and the Josephson regime.
Adopting the reflection principle and the Poisson summation 
formula we derive an analytical expression for the dynamics of the population imbalance depending 
on the few relevant parameters of the system only. This allows us to discuss 
its characteristic dynamics, especially the oscillation frequency, 
and the collapse- and revival time, as a function of the model parameters, leading 
to a deeper understanding of Josephson physics. We find that our fomulae match
previous experimental observations.

\end{abstract}

\maketitle

\section{Introduction}
Fundamental issues of non-equilibrium physics of interacting
many-body quantum systems and of phase coherence and phase stability, in
particular, have a long history. A simple yet relevant model, the two-site Bose-Hubbard
Hamiltonian, features phase and fluctuation decay, and also revivals and thus,
over the years, many thorough investigations of its quantum dynamics have appeared.
Most remarkably, recent experiments involving ultracold Bose gases trapped in
an effectively one-dimensional double-minimum potential represent an almost ideal
realizations of this fundamental model \cite{Alb05,Gat07}, with the fascinating
possibility to vary relevant model parameters over a wide range.

A full many-body calculation of the dynamics of an interacting, trapped ultracold Bose gas is only 
possible for a very small number of particles, even for weakly interacting Bosons.
Most often a mean field approximation in form of the Gross-Pitaevskii equation
is applied, which provides good results for low temperatures and for a large number
of particles $N$, if only for a limited time and a limited set of observables.
These limits are intensively studied. Once the field operators are replaced by a c-number field, 
some truly quantum phenomena, e.g. wave function revivals, cannot be described.
The double-well potential provides an ideal playground to analyze these issues.
Thus, a purely classical field approach quickly comes to its limits, and the question
arises whether {\it semiclassical} methods can improve the
theoretical treatment of such bosonic systems, allowing us in the future to
study more challenging problems whose many-body Schrödinger equation can no longer 
be solved fully numerically. 

A number of articles deal with the discussion of the consequences of the 
mean-field approximation and many-body quantum corrections 
\cite{Var01,Ang01}
  and the many-body quantum and classical dynamics in phase space 
\cite{Mah05}.
Furthermore, semiclassical methods were applied to the double-well system. In
\cite{Gra07,Paw11,Fran00,Chu10,Sh08,Nis10} a WKB quantization is adopted to
analyze the energy spectrum and the wave functions in certain parameter regions.

Despite this fair amount of investigations, it is remarkable to realize that
-- leaving some fairly straightforward cases aside --
no analytical expressions for the relevant dynamical quantities
appear to be known. Thus,
the purpose of this article is to find a generally applicable analytical
description of the 
population imbalance dynamics of an ultracold Bose gas in a double-well potential 
by applying semiclassical methods. Since the full quantum dynamics can be determined numerically 
up to many thousand particles, we are able to compare to exact results.
Clearly, the interesting case of very large $N 
\rightarrow \infty$ can no longer
be investigated numerically, yet our analytical approach is suited to study this
very limit in detail.

At low temperatures a Bose-Einstein condensate in a double-well 
potential can be described by a two-mode approximation.
The corresponding second quantized many particle two-site
Bose-Hubbard Hamiltonian is written as
\begin{equation}
  \hat{H}_{BH}=-T\left( \hat a_1^\dagger \hat a_2 + \hat a_2^\dagger 
\hat a_1 \right) + U\left(\hat a_1^\dagger \hat a_1^\dagger \hat a_1 
\hat a_1
+ \hat a_2^\dagger \hat a_2^\dagger \hat a_2 \hat a_2 \right) + \delta 
\left( \hat n_1 - \hat n_2 \right)
\label{eqBH}
\end{equation}

with the creation and annihilation operators for a boson in the $i$th 
well denoted by $\hat a_i^\dagger $, $\hat a_i$ with $[a_i,a^\dagger_j]=\delta_{ij}$. Thus,
the particle number operator of the $i$th site is $\hat n_i = \hat 
a_i^\dagger \hat a_i$. $U$ is a measure for the on-site two-body 
interaction strength, $T$
is a tunneling amplitude, which in the experiments can be controlled by
varying the barrier hight. The tilt parameter 
$\delta$ leads to an asymmetry in the one-particle site
energies of the two wells and is used to initiate the dynamics.
Note that in the standard notation adapted in Josephson physics we have
$E_J=NT$ and $E_C=4U$ \cite{Leg01}.
\\
It has been shown that the Bose-Hubbard Hamiltonian describes the 
dynamics of the bosons in the double-well potential properly 
\cite{Mil97}, provided that the
interaction energy $U$ is small compared to the level spacing  of the 
trap potential, such that only the two lowest lying modes have to be 
taken into account. Transverse modes should also be suppressed.
It should be mentioned that there are finer descriptions of the two-mode
limit that also take into account tunnel coupling energies depending 
explicitly
  on the nonlinear two-body interaction term \cite{An06}. In this work,
  however, we restrict ourselves to the standard Bose-Hubbard
  Hamiltonian (\ref{eqBH}). 
\\
\\
First, there are three qualitatively quite different regimes \cite{Pa01,Leg01}
with respect to crucial features of
the energy spectra. They are best explained by introducing the parameter
\begin{equation}
\Lambda=\frac{U N}{T},
\end{equation}
which thus separates the Rabi- ($\Lambda < 1$) from the so-called
Josephson regime for which $1<\Lambda\ll N^2$ and the Fock regime with $\Lambda \gg N^2$. 

The Rabi regime is the 
non-interacting limit $\Lambda \ll 1$, when the system consists of $N$ 
independent particles leading to an almost harmonic oscillator energy
spectrum and thus, after an initial tilt, to plasma oscillations with the
known plasma frequency
$\omega_p=2T\sqrt{1+\Lambda}\approx 2T$ \cite{Smer97,Gat07}.

In the Fock regime all eigenenergies are grouped in 
doublets with a quasi-degenerate symmetric and antisymmetric state. Thus, the dynamics 
of the mean population imbalance
follows an extremely slow evolution in time which is called self trapping. 

The Josepshon regime combines the two characteristics of the spectrum just discussed. 
We distinguish the
self trapping regime $E>2NT$ from the plasma oscillating regime,
where $E < 2NT$ holds. In the former, the energy eigenstates 
appear as doublets again leading to self trapping.
In the latter the energy eigenstates correspond to an (an-harmonic) oscillator 
spectrum and the population imbalance oscillates around zero. 

Thus, in the Josephson regime the dynamics will depend on the energy of the 
initial state. For low energies -- the subject of this work -- the dynamics undergoes
plasma oscillations, for higher
energies we see self-trapping, which is beyond the scope of this paper.
\\
\\
In this article we have in mind an experiment as in reference 
\cite{Alb05}, so the double-well system is initially prepared in the 
ground state $\psi_0$ of
  a tilted potential, i.e. $\delta \ne 0$ in (\ref{eqBH}). Then, at $t=0$ it is 
quickly switched to a symmetric potential, i.e. $\delta = 0$.
Starting from an initial population imbalance unequal to zero the system is 
left to evolve in time.
\\

In our paper we first discuss the spectrum using the semiclassical WKB-
or Bohr-Sommerfeld quantization. We find a way to systematically obtain
an approximate, useful expression for energies in the plasma oscillating
regime. In order to describe imbalance dynamics, we need to explore
overlap matrix elements in the following section, which we do with the help
of the reflection principle. We then apply the Poisson summation
formula, which has a long history in semiclassical approaches to
quantum dynamics. As a result, we find a useful expression for the
time evolution of the imbalance, containing parameters that can be
obtained analytically on the basis of the classical Hamiltonian.
We then compare exact calculations with our new formula and find
remarkable agreement over the whole relevant range of $\Lambda$,
covering the known Rabi- but also the plasma oscillating Josephson
region. In particular, the oscillation frequency, the collapse and
revival times are reproduced astonishingly well. We finally discuss the
corresponding analytical expressions. It should be noted that the
experimentally observed oscillation frequency in \cite{Alb05} of about
$40$ms follows directly from our formula.

\section{Semiclassical description}
\label{sec_semicl}
We will follow mainly Braun \cite{Bra93} and his discrete WKB method,
as already applied to the double-well problem by Korsch et 
al. \cite{Gra07}. The two-mode Bose-Hubbard Hamiltonian can be written in the 
Schwinger spin
representation by transforming to angular momentum 
operators  $\hat J_x=\frac{1}{2}\left(\hat a_1^\dagger \hat a_2 + \hat 
a_2^\dagger \hat a_1 \right)$,
  $\hat J_y=\frac{1}{2i}\left(\hat a_1^\dagger \hat a_2 - \hat 
a_2^\dagger \hat a_1 \right)$ and
$\hat J_z=\frac{1}{2}\left(\hat a_1^\dagger \hat a_1 - \hat a_2^\dagger 
\hat a_2 \right)$. With the ladder operators $\hat J_+=\hat J_x+i 
\hat J_y$ and
$\hat J_-=\hat J_x - i \hat J_y$ 
the Hamiltonian (\ref{eqBH}) becomes
\begin{equation}
  \hat H =2U \hat J_z^2+2 \delta \hat J_z -T\left(\hat J_+ + \hat J_- 
\right) + \frac{1}{2} U \hat N^2 - U\hat N \ ,
\end{equation}
where $\hat N$ is the total particle number operator. For fixed $N$ a
change from basis $|n,N-n\rangle$ to the angular momentum states $|l,j\rangle$
is useful, with 
$l=\frac{N}{2}$ and $j=\frac{n_1-n_2}{2}$.
With $w_j=2Ul^2-2Ul+2Uj^2+2 \delta 
j$ and $p_j=-T\sqrt{l(l+1)-j(j-1)}$, the eigenvalues of the Hamiltonian are
determined by an equation of the form
\begin{equation}
p_j c_{j-1} + (w_j - E) c_j + p_{j+1} c_{j+1}=0 \ ,
\label{eqSG}
\end{equation}
as discussed in \cite{Bra93}.
By introducing the ``coordinate'' operator
 $\phi=i 
\frac{\partial}{\partial j}$ (note \cite{BraunFootnote}),
equation (\ref{eqSG}) can be written as a 
Schr\"odinger equation for the function $c_j$ with
eigenvalue $E$ and Hamilton operator $\hat H = w(j)+p(j)e^{-i \phi} + 
p(j+1) e^{i \phi}$. In the classical limit the operators 
turn to 
canonically conjugate coordinate $\phi$ and momentum $j$ (population imbalance), where 
$\phi$ turns out to be the phase difference 
between the two wells.
Since $p(j)$ is a slowly varying function of $j$ in the classical 
limit ($N \rightarrow \infty$) one can replace both $p_j$ and $p_{j+1}$ by $p_{j+\frac{1}{2}}$ and 
one finds the Hamilton function
\begin{align}
  H(j,\phi) &= w(j)+2p(j+\frac{1}{2})\cos \phi \\
&= \frac{1}{2} U N^2-UN+2Uj^2+2\delta j -2T\sqrt{(N/2)^2-j^2}\cos \phi
\label{eqHamiltonfunktion}
\end{align} 
which can also be found from the mean-field
Gross-Pitaevskii functional in the two-mode limit \cite{Smer97, NFootnote}.
The classical dynamics of the population imbalance and the relative phase (for 
$\delta=0$) is then determined by Hamilton's equations of 
motion:
\begin{align}
  \frac{dj}{dt}&=-\frac{\partial H}{\partial \phi}=-2T \sqrt{(N/2)^2-j^2} 
\sin \phi
\nonumber
\\
\frac{d \phi}{dt}&=\frac{\partial H}{\partial j}=4Uj+\frac{2Tj \cos 
\phi}{\sqrt{(N/2)^2-j^2}}
\label{eqofmotion}
\end{align}

The rich dynamics in this ''classical picture'' have been studied by 
several groups \cite{Mil97, Rag99, Hol01}, focusing on 
the differences between the classical and the quantum description of the 
dynamics \cite{Kra09,Jav10,Ton05}. Clearly a purely classical description 
cannot picture the collapses and the revivals of the population imbalance,
but it is able to shed light on the transition from the tunneling to the 
self-trapping regime.
Recently, the phase space region near the classical bifurcation was also investigated 
experimentally with ultracold Bosons \cite{Zib10}.

\subsection{Semiclassical energy spectrum: Bohr-Sommerfeld quantization}
An analytical approach to the energy spectrum relies on the WKB method 
following Braun \cite{Bra93} and others \cite{Fran00,Chu10,Gra07,Paw11}. 
In \cite{Gra07}, only the noninteracting case is investigated analytically. In \cite{Fran00,Chu10} 
 the authors concentrate on energies close to the extremal points,
 and in \cite{Paw11} the case of an attractive gas for the single value of $\Lambda=1$ 
is studied. We here concentrate on the plasma oscillating regime and aim 
 for solutions over the whole range of $\Lambda \ll 1$ to $\Lambda \gg 1$. 
 
\begin{figure}[htb]
\includegraphics[width=40mm]{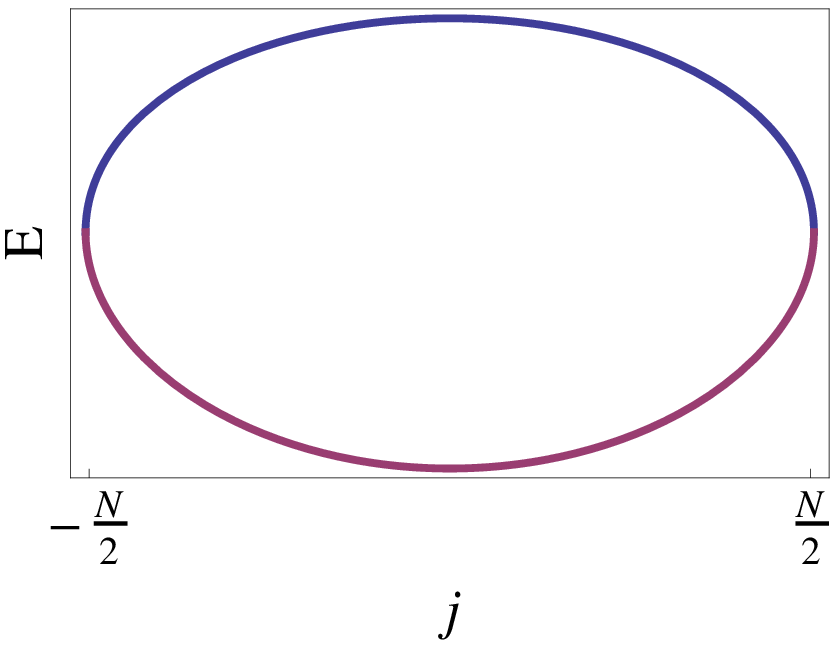} \hspace{1cm}
\includegraphics[width=40mm]{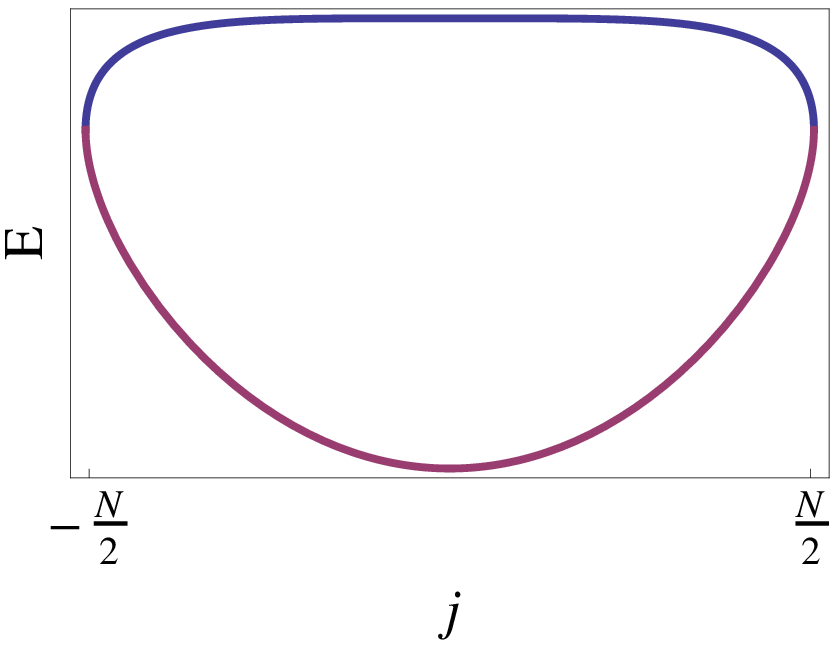} \hspace{1cm}
\includegraphics[width=40mm]{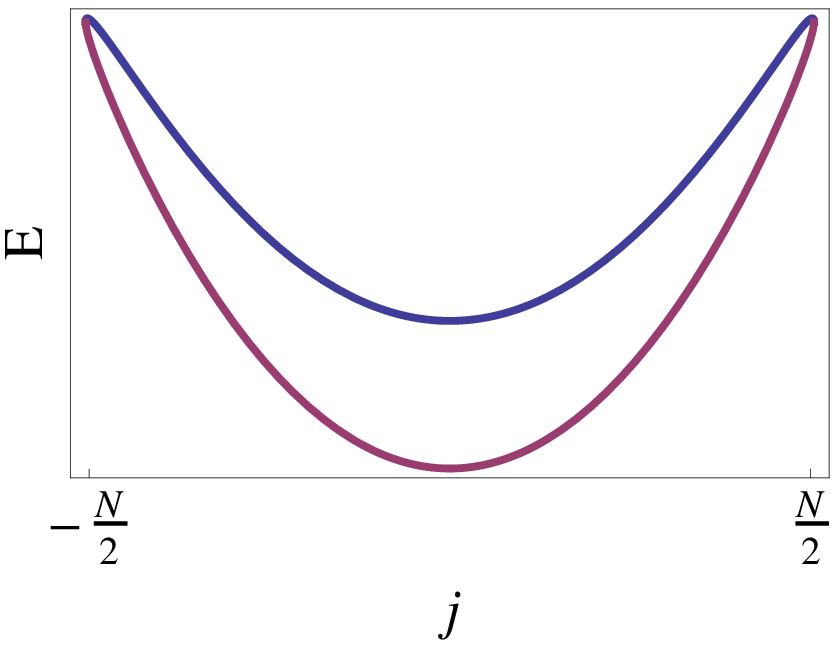}
\\
\includegraphics[width=40mm]{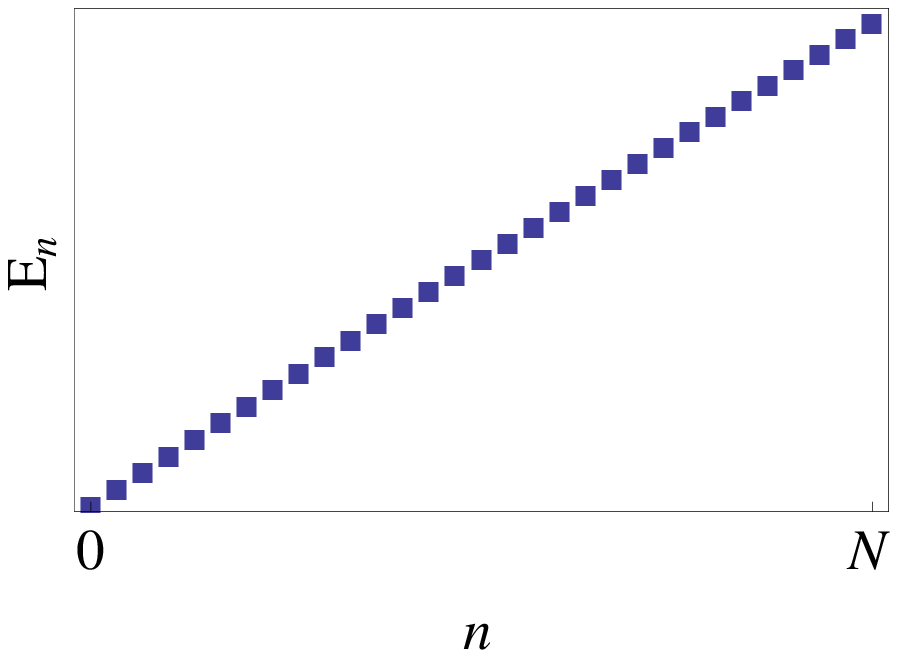} \hspace{1cm}
\includegraphics[width=40mm]{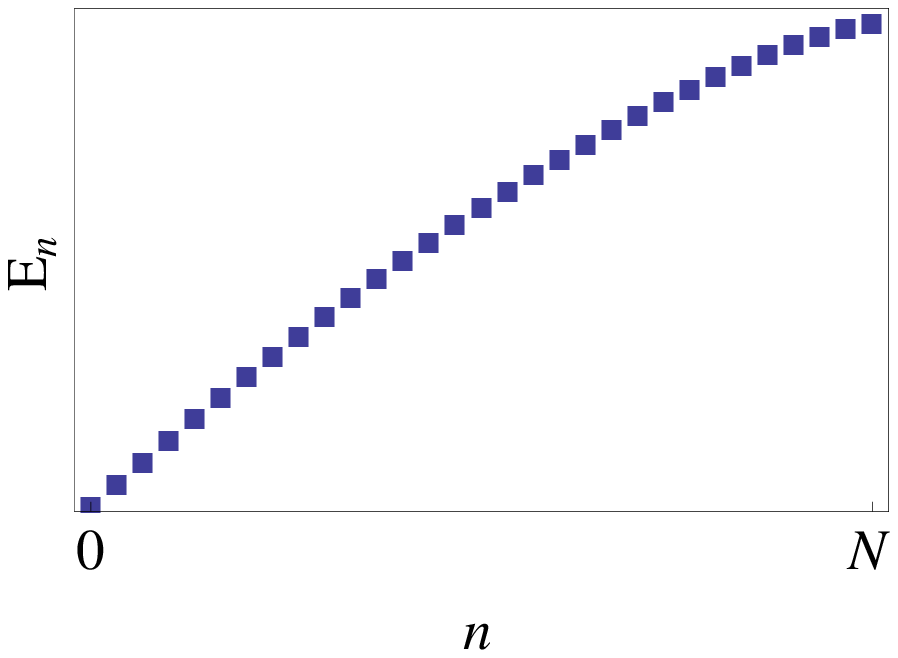} \hspace{1cm}
\includegraphics[width=40mm]{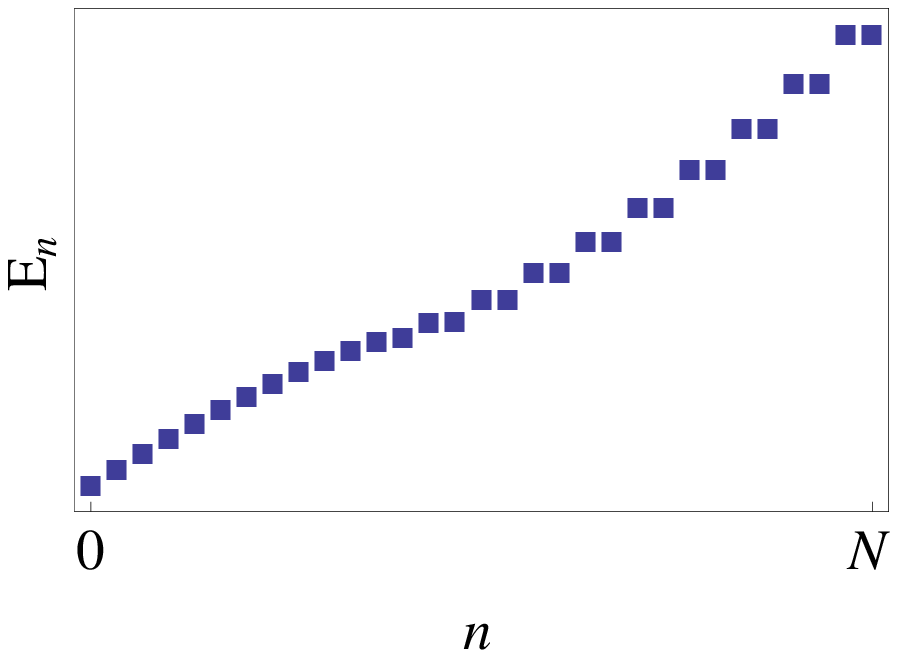}
\caption{Potential curves $V^+$, $V^-$ and the energy eigenvalues $E_n$ for, from left to right, $\Lambda=$ 0.1, 1, 10 and $N=30$, $T=3$.  }
\label{figPotentials}
\end{figure}

For the Hamilton function (\ref{eqHamiltonfunktion}) it is 
convenient to introduce two potential-energy curves
\begin{align}
  V^+(j)=H(j,\pi)=TN+2Uj^2+2T\sqrt{(N/2)^2-j^2}
\\
V^-(j)=H(j,0)=TN+2Uj^2-2T\sqrt{(N/2)^2-j^2}
\label{eqpotentials}
\end{align}
such that the classically allowed energies lie in the region confined by 
the two potential curves $V^+$ and $V^-$. The minimum energy is chosen to be $V^-(j=0)=0$. The potential curves display 
the transition from the Rabi- to the Josephson
  regime very nicely, as shown in Fig. \ref{figPotentials}.
The energy eigenvalues change from a (non-harmonic) oscillator like spectrum for $\Lambda < 1$ to a 
spectrum with doublets for $\Lambda > 1$ due to tunneling,
which can be seen from the potential curves. For $\Lambda > 1$, $V^+$ attains a local 
minimum which leads to doublets in the spectrum for energies $E>V^+(0)$. The deeper 
the minimum, the bigger this so-called Fock-fraction of the spectrum. Since 
we are interested in plasma oscillations, the Fock-fraction will not be 
 investigated here, but a semiclassical analysis along similar lines
 -- if only more involved -- is possible, see for instance \cite{SAB91}.
\\
\\
In the WKB approximation the eigenenergies $E_n$ are obtained from the quantization condition 
\begin{equation}
S=S(E)= \oint \phi(j) dj = 4\int_{0}^{j_+(E)} 
\arccos\left(\frac{E-TN-2Uj^2}{2T\sqrt{(N/2)^2-j^2}}\right)dj=2\pi(n+\frac{1}{2}) 
\ ,
\label{eqWKB}
\end{equation}
where $n$ is the quantum number, and $\phi(j)$ is determined by the 
Hamilton function (\ref{eqHamiltonfunktion}) at fixed energy $E$ 
(recall that zero energy $E=0$ corresponds to $S(E=0)=0$). The integration limit $j_+$ is 
the (positive) classical turning point as obtained from
\begin{equation}
  \sqrt{(V^+(j)-E)(E-V^-(j))}\stackrel{\mathrm{!}}=0,
\end{equation}
which leads to a quadratic equation in $j^2$ with solutions
\begin{equation}
(j^2)_\pm(E) = \frac{1}{2U^2}\left( 
(EU-(\omega_p/2)^2)\pm\sqrt{(\omega_p/2)^4-EU\omega_p^2/(2(1+\Lambda))}\right).
\end{equation}
Recall that $\omega_p=2T\sqrt{1+\Lambda}$ is the plasma frequency. For the plasma oscillating regime the relevant turning point is $j_+$. 
Note that $j_+\rightarrow 0$ for $E\rightarrow 0$, while $(j_-^2)$ 
approaches the negative constant $(j_-^2)\rightarrow - 
(\omega_p/(2U))^2$ as $E\rightarrow 0$.

The integral in eqn. (\ref{eqWKB}) can be solved numerically and the results 
agree very well with the exact quantum results even for quite small numbers 
of particles as has already been noticed in \cite{Gra07}.
It is impossible to 
solve the action integral analytically without approximation. As we aim 
at the plasma oscillation regime, we expand in powers of $E$. First,
however,
we take the derivative with respect to energy and rescale to find
\begin{equation}
  \frac{\partial S(E)}{\partial E}=\frac{2}{U|j_-(E)|}\int_{0}^1 
\frac{d\lambda}{\sqrt{(1-\lambda^2)(1+\kappa^2(E)\lambda^2)}} \ .
\label{eq_integral}
\end{equation}
with $\kappa^2=(j_+)^2/|j_-|^2$.
Since $\kappa^2 \rightarrow 0$ for $E\rightarrow 0$, and $0<\lambda<1$, an expansion of $(1+\kappa^2\lambda^2)^{-1/2}$ in powers of $\kappa^2\lambda^2$ 
leads to a series in powers of $E$. The corresponding integrals $\int_0^1 \mathrm{d}\lambda \frac{\lambda^{2n}}{\sqrt{1-\lambda^2}}$ are
 known analytically. Finally, a systematic expansion of $\kappa^{2n}$ and $1/|j_-|$ in $E$ leads to
\begin{equation}
\frac{\partial S}{\partial E} = \frac{2 \pi}{\omega_p}+4\pi 
\frac{U(1+\Lambda/4)}{\omega_p^3 (1+\Lambda)}E + 6\pi 
\frac{3U^2(1+\Lambda/3+(\Lambda/4)^2)}{\omega_p^5(1+\Lambda)^2}E^2 + ....,
\label{eq_sstrich}
\end{equation}
which is one of the important results of this paper.
Apparently, the formal expansion in $E$ is an expansion in the dimensionless parameter 
\begin{equation}
 \varepsilon = \frac{UE}{\omega_p^2}=\frac{1}{2}\frac{\Lambda}{1+\Lambda}\left(\frac{E}{V^+(0)}\right) \ .
\end{equation}
The expression on the right hand side clearly shows that our results are expected to be valid in 
 the plasma oscillating regime $E < V^+(0)$, irrespectively of the value of $\Lambda$. 
From a simple integration together with the Bohr-Sommerfeld-quantization condition (\ref{eqWKB}) 
we find
\begin{equation}
  n(E)=-\frac{1}{2}+\frac{1}{\omega_p}E + 
\frac{U(1+\Lambda/4)}{\omega_p^3 
(1+\Lambda)}E^2+\frac{3U^2(1+\Lambda/3+(\Lambda/4)^2)}{\omega_p^5(1+\Lambda)^2}E^3 
+ ...
\label{eqspectrum}
\end{equation}

\begin{figure}[htb]
\includegraphics[width=50mm]{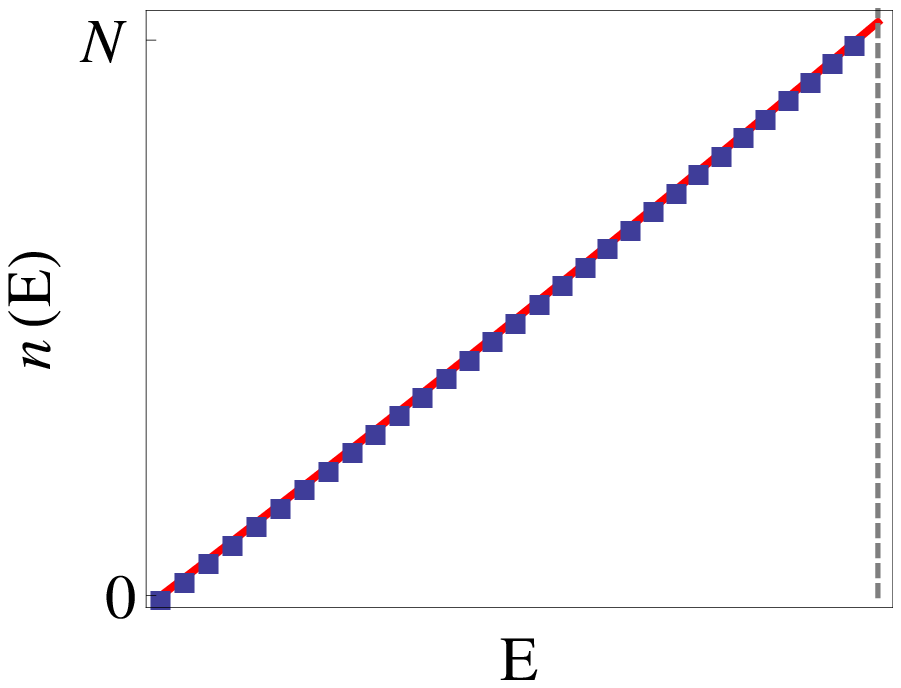}
\includegraphics[width=50mm]{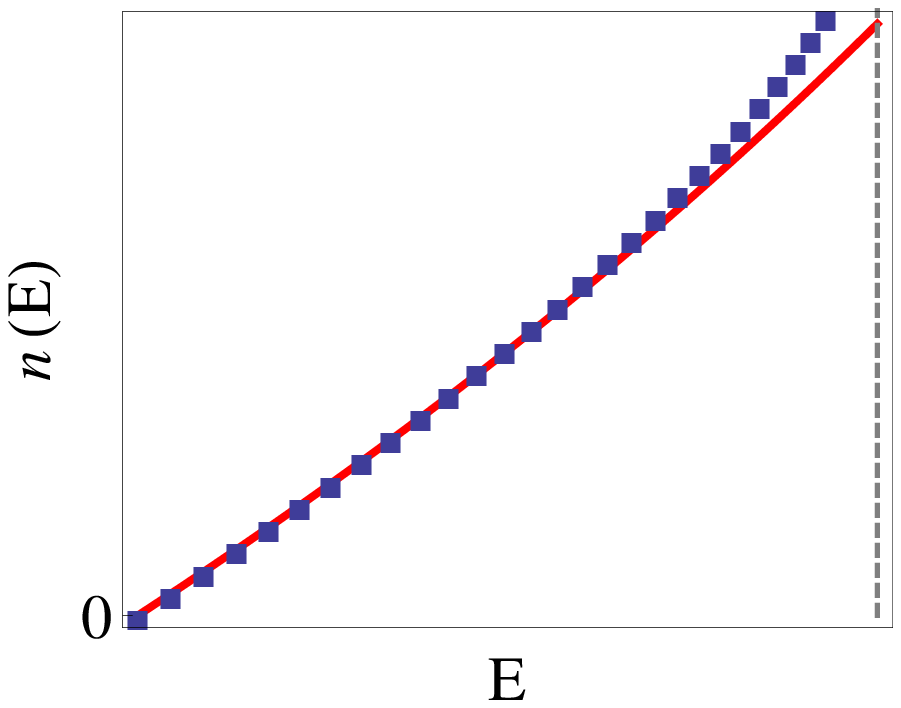} 
\includegraphics[width=50mm]{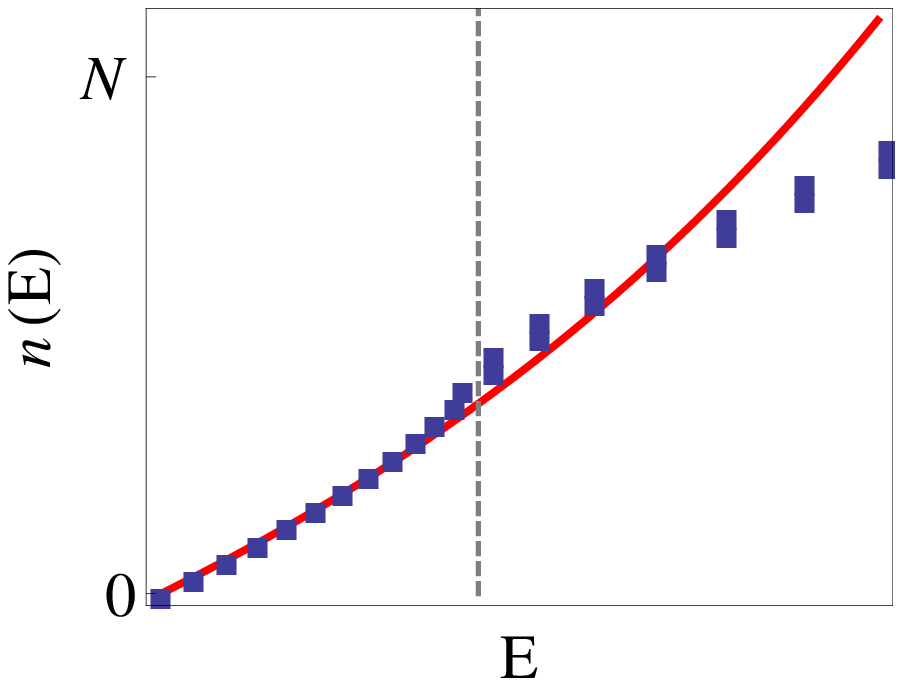}
\caption{Comparison of the analytical (\ref{eqspectrum}) (red, solid 
line, including third order in $E$) and the numerically exact spectrum (blue squares)  for,
from left to right, $\Lambda=$ 
0.1, 1, 10 and for $N=30$, $T=3$. The vertical dashed lines illustrate the transition from the plasma oscillating regime to the self trapping regime at $E=V^+(0)=2NT$. We see excellent agreement in the plasma oscillating regime.  }
\label{figspectrum}
\end{figure}
In figure \ref{figspectrum} we show examples of the spectrum for a wide range of values of $\Lambda=0.1, \ 1, \ 10$, 
covering both the Rabi and the Josephson regime. Apparently, our approximation 
(\ref{eqspectrum}), including contributions up to third order in $E$, 
coincides with the numerically exact spectrum with high 
accuracy in the plasma oscillating regime ($E<V^+(0)$) for all values of $\Lambda$.
Clearly, 
the doublet structure in the Fock regime (high energy regime $E>V^+(0)$ in the right diagram of Fig. \ref{figspectrum})
cannot be captured by our series expansion (\ref{eqspectrum}).

\section{Exact quantum dynamics of the population imbalance}

To determine the tunneling dynamics, the Bose-Hubbard Hamiltonian 
(\ref{eqBH}) can be diagonalized numerically for a finite number of Bosons. 
Using the eigenbasis 
$\{|\phi_n\rangle\} $, the dynamics of $|\psi(t) \rangle $ is given by
\begin{equation}
  |\psi(t)\rangle = \sum _n c_n e^{-iE_nt} |\phi_n\rangle \ ; \ \text{with} \ c_n=\langle \phi_n | \psi_0\rangle
\end{equation}
The time evolution of the population imbalance $\hat j=(\hat n_1-\hat n_2)/2$ is then
\begin{equation}
  j(t)=\langle \psi(t) | \hat j |\psi(t)\rangle=\sum_{n,m} 
A_{nm}e^{-i(E_n-E_m)t}
\label{eqImbalance}
\end{equation}
with the matrix
\begin{equation}
  A_{nm}=c_n c_m^\ast \langle\phi_m|\hat j|\phi_n\rangle \ .
\label{eqAnm}
\end{equation}
The dynamics of the 
population imbalance thus depends on the energy spectrum through the differences $E_n-E_m$, 
and on the matrix 
$A_{nm}$, which contains the initial condition and matrix elements 
$\langle\phi_m|\hat j|\phi_n\rangle$.

\begin{figure}[htb]
\includegraphics[width=30mm]{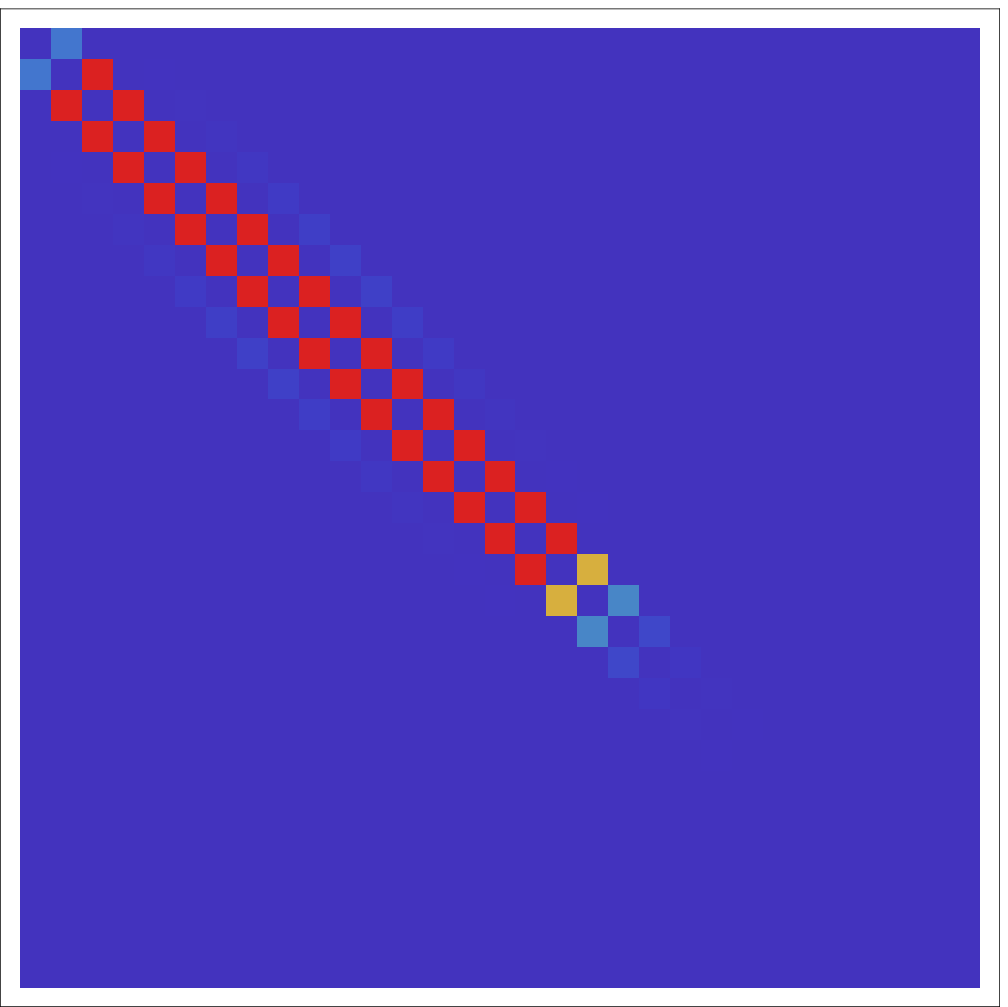} \hspace{1cm}
\includegraphics[width=30mm]{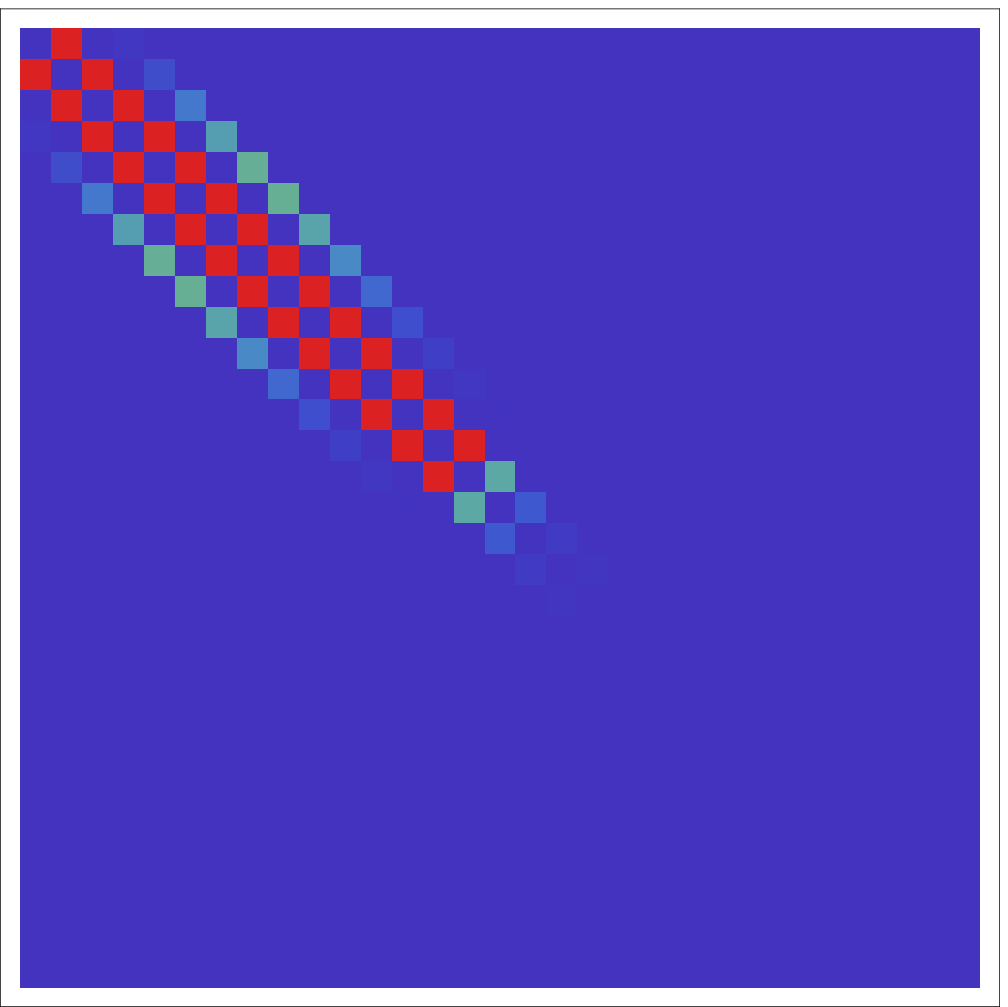} \hspace{1cm}
\includegraphics[width=30mm]{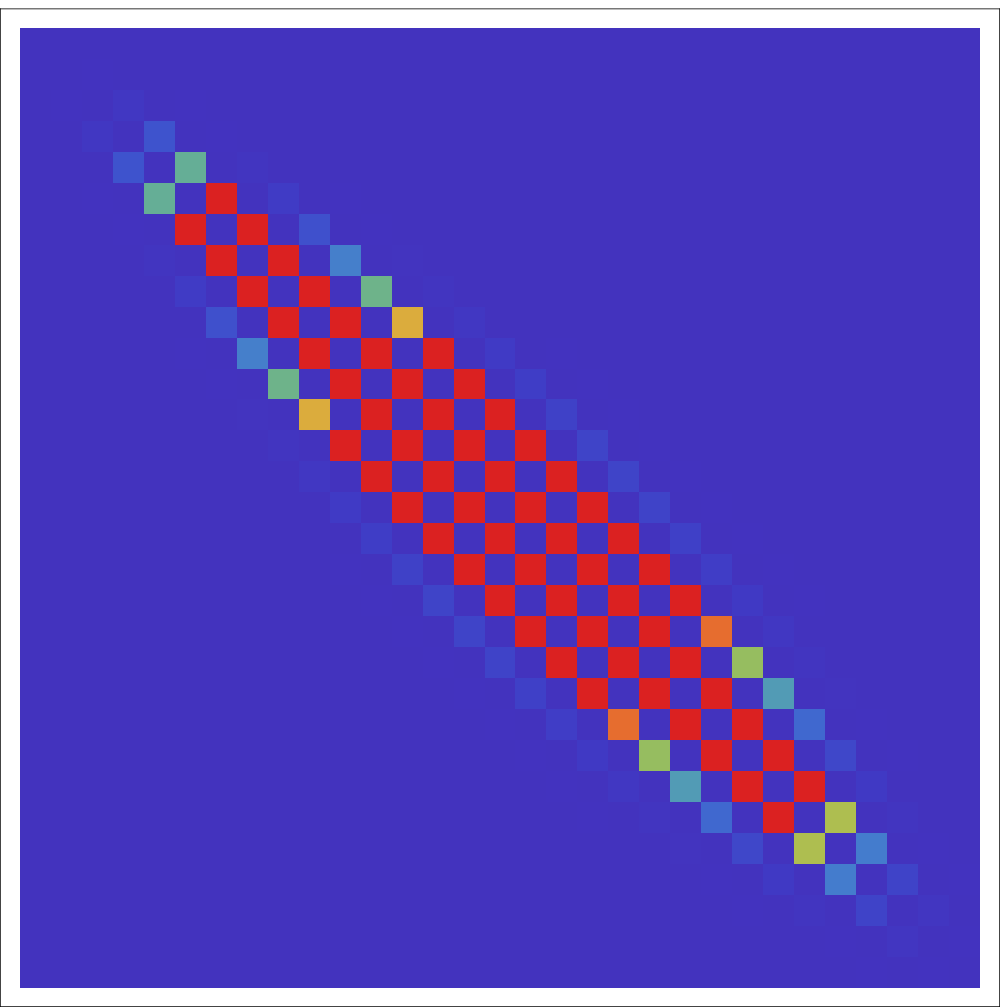}\hspace{1cm}
\includegraphics[width=37mm]{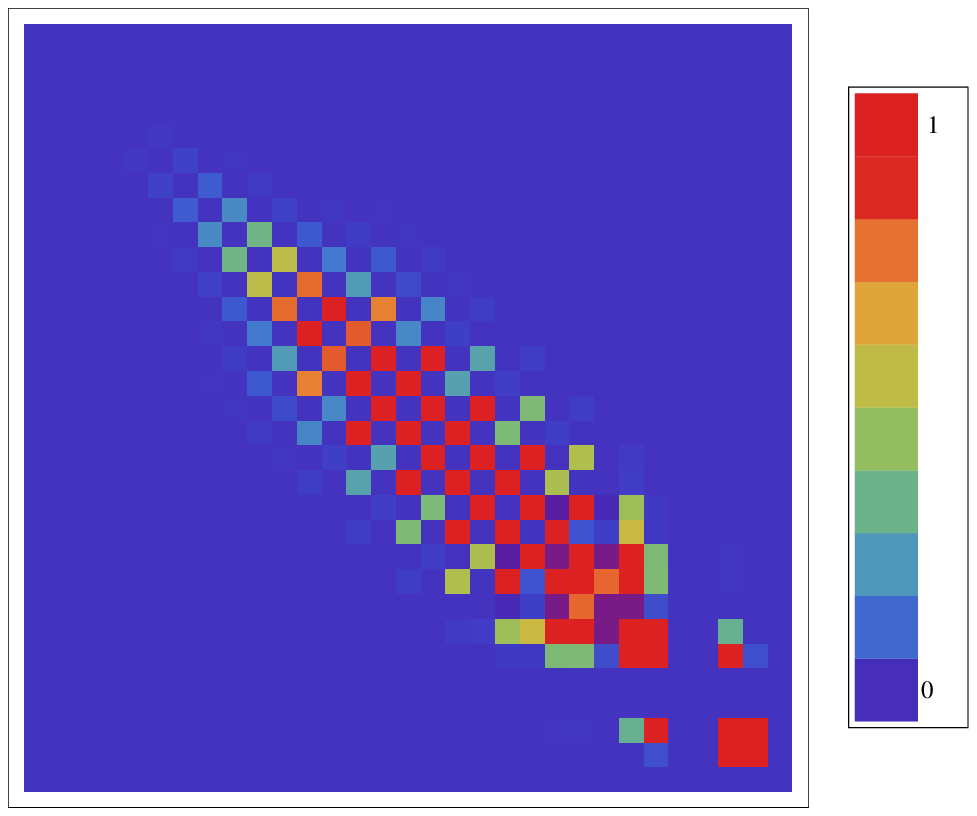}
\caption{The matrix $A_{nm}$  for $\Lambda=$ 0.1, 0.5, 1, 10 and for 
$N=30$, $T=3$.  }
\label{figAnm}
\end{figure}

Figure \ref{figAnm} shows the matrix $A_{nm}$ for increasing $\Lambda$,
obtained from a numerically exact calculation. 

Due to parity with respect to $\hat j$, $A_{nm}$ is zero for an even number $n-m$, 
as can be seen in Fig. \ref{figAnm}: $0=A_{nn}=A_{nn\pm 2}=A_{nn\pm 4}+\ldots$. 
Among the non-zero matrix elements, there is a strong hierarchy, 
\begin{equation}
 |A_{nn\pm 1}|\gg |A_{nn\pm 3}|\gg |A_{nn\pm 5}| \gg \ldots \ ,
 \label{eq_hierarchy}
\end{equation}
in particular for small $\Lambda$, which will be important later. It
 is worth noting that in the limit $U\rightarrow 0$ (and therefore $\Lambda \rightarrow 0$,
 for fixed $N$) 
 the dynamics is well described by a harmonic oscillator. In that case  
it is easy to prove (the $\phi_n(j)$ are Hermite polynomials) that only the $A_{nn\pm 1}$ are
in fact different from zero.

Along the diagonals, the matrix elements $A_{nn\pm k}$ 
(with $k=1, \ 3, \ 5, \ \ldots$) have a Gaussian-like $n$-dependence.  
This is due to the $n$-dependence of the overlap $c_n=\langle \phi_n | \psi_0\rangle$, 
 which will be discussed in the next section. 
By contrast, the $n$-dependence of the matrix elements $\langle\phi_{n\pm k}|\hat j|\phi_{n}\rangle$
 is weak. Thus, it is save to assume the form
 \begin{equation}
  A_{nn\pm k} \approx c_n c_{n\pm k}^\ast d_k \ ,
  \label{eq_dk}
 \end{equation}
with $n$-independent parameters $d_k\approx \langle \phi_{\bar n \pm k}|\hat j|\phi_{\bar n}\rangle$ 
(with the most relevant ${\bar n}$), for which, following (\ref{eq_hierarchy}), we expect 
\begin{equation}
 |d_1| \gg |d_3| \gg |d_5| \gg \ldots 
 \label{eq_hier_dk}
\end{equation}

\section{Semiclassical dynamics of the population imbalance}
For a semiclassical evaluation of $j(t)$ according to equations (\ref{eqImbalance}) and (\ref{eqAnm}) we 
need semiclassical expressions for $E_n-E_{n\pm k}$ and the overlap coefficients $c_n$. While 
the spectrum was discussed in section \ref{sec_semicl}, we start here with the latter. 

\subsection{Reflection principle}
The problem to find overlap integrals of an initial wavepacket $\psi_0(j)$ 
localized near $j \approx j_0$
with eigenstates $|\phi_E\rangle$ of the Hamiltonian with potential $V(j)=V^-(j)$
is often encountered in molecular photo-dissociation \cite{schinke}. The semiclassical 
solution (reflection principle) states that 
\begin{equation}
 \langle \phi_E|\psi_0\rangle = c \cdot \psi_0\left(\frac{E-V(j_0)}{V'(j_0)}\right) \ 
\end{equation}
with some constant $c$.
It is important to note that here the eigenstates are understood to be energy normalized, 
i.e. $\langle \phi_E|\phi_{E'}\rangle=\delta(E-E')$, since in typical applications these
are scattering states. For the coefficients
$c_n$ we therefore find $c_n=\langle \phi_n|\psi_0\rangle=\sqrt{\frac{\mathrm{d}E}{\mathrm{d}n}}\langle \phi_E|\psi_0\rangle$.
The normalization condition $1=\sum_n |c_n|^2 \approx \int \mathrm{d}n |c_n|^2$ yields $c=1/\sqrt{V'(j_0)}$, and we get 
\begin{equation}
c_n=\langle\phi_n|\psi_0\rangle \approx
\frac{1}{\sqrt{V'(j_0)}}\sqrt{\frac{\mathrm{d}E(n)}{\mathrm{d}n}} \; \psi_0\left(\frac{E_n-V(j_0)}{V'(j_0)}\right) \ .
\label{eq_cninE}
\end{equation}
In our calculations, following the experiments, the initial wave function is 
prepared as the ground state of the tilted trap potential (achieved through the term $\delta (\hat n_1-\hat n_2)=2\delta \hat j$ in the Bose-Hubbard 
Hamiltonian (\ref{eqBH})). 
In a harmonic approximation near the potential minimum of the tilted potential we find the Gaussian density
\begin{equation}
  |\psi_0(j)|^2=\frac{1}{\sqrt{2 \pi \sigma^2}}e^{-\frac{(j-j_0)^2}{2 
\sigma^2}}
\label{eqpsi0}
\end{equation}
with $j_0$ uniquely determined by the tilting strength $\delta$ and
\begin{equation}
\sigma^2=\frac{N}{4} 
\frac{1-(2j_0/N)^2}{\sqrt{1+\Lambda(1-(2j_0/N)^2)^{3/2}}} \ .
\end{equation}

Clearly, the shape of the initial wave function determines the shape of the $c_n$ as a function of $n$. 
On closer inspection of equation (\ref{eq_cninE}), however, we observe that for the initial state (\ref{eqpsi0}), due to the nonlinear relation between $E_n$ and $n$, 
the coefficients $c_n$ are gaussian in $E_n$ but not in $n$. 

\subsection{Population imbalance}\label{subsecpop}
Having all the ingredients at hand we can now aim at a semiclassical expression 
for the dynamics of the population imbalance $j(t)$ which we choose to 
write as 
\begin{equation}
  j(t)= \sum_{n,k} A_{nn-k} \exp(-i (E_n-E_{n-k})t) + c.c.
\label{eqdynamics}
\end{equation}
with $k=1,3,5,\ldots$ taking into account the diagonal structure of $A_{nm}$ as discussed in the last section. 
Replacing $A_{nn\pm k}$ by expression (\ref{eq_dk}) and using the Poisson summation formula we 
find 
\begin{equation}
 j(t)=\sum_{k=1,3,5..} d_k \sum_{m=-\infty}^\infty I_m^k(t) + c.c.
 \label{eq_j_t}
\end{equation}
with 
\begin{equation}
 I_m^k(t)=\int \mathrm{d}n 
 \left(\frac{\mathrm{d}E}{\mathrm{d}n}\right)
 \frac{1}{V'(j_0)} 
 \psi_0\left(\frac{E_n-V(j_0)}{V'(j_0)}\right) 
\psi_0^\ast\left(\frac{E_{n-k}-V(j_0)}{V'(j_0)}\right) 
e^{-i(E_n-E_{n-k})t} e^{2\pi i mn} \ .
\end{equation}
This rather complicated expression is readily simplified by changing the 
integration variable from $n$ to $E$. Further, as only very small $k$ ($k=1, \ 3)$ are 
relevant (see equ.(\ref{eq_hier_dk})), it is safe to replace $E_n-E_{n-k} \approx \frac{\mathrm{d}E}{\mathrm{d}n}k= \frac{2\pi k}{S'(E)}$ 
and neglect the $k$-dependence in the reflection principle, i.e. $c^\ast_{n\pm k} \approx c^\ast_n$. 
Finally, we replace $2\pi n=S(E)-\pi$ according to the semiclassical quantization rule (\ref{eqWKB}). 
With $\tau=kt$ we find

\begin{equation}
I_m^k(t)=I_m(\tau)= e^{i\pi m} \int \frac{\mathrm{d}E}{V'(j_0)} \bigg|\psi_0 
\left(\frac{E-V(j_0)}{V'(j_0)} \right)\bigg |^2 
e^{-\frac{2 \pi i \tau }{S'(E)}}e^{i m S(E)} .
\label{eq_Imk}
\end{equation}
This expression, together with equ.(\ref{eq_j_t}) is one of the main results of our paper. 
As we will see, even with further simplifications, the formula captures all essential details
of the dynamics, 
allows for a thorough understanding of decay and revival dynamics, and, 
most importantly, is the starting point for analytical expressions. 
\\
\\
Due to the localization of the initial state $\psi_0(j)$, the
energy integration in (\ref{eq_Imk}) is confined to a relatively small interval near $E\approx V(j_0)$, 
which we assume to be in the plasma oscillating regime ($E<V^+(0)$). Therefore, for the 
evaluation of 
the overall phase $mS(E)-2 \pi kt /S'(E)$ we can rely on our 
semiclassical series expansions (\ref{eq_sstrich}) and (\ref{eqspectrum}). 
With a Gaussian initial state as in (\ref{eqpsi0}) and expanding the  
overall phase up to second order around $E\approx V(j_0)$ allows us 
to take the Gaussian integral and leads us to the analytical result

\begin{equation}
  I_m(\tau)+c.c.=\frac{2}{(1+A^2)^{1/4}}\cos\left(\tilde{\omega}_p \tau -\tilde \varphi \right)
    \exp\left(-\frac{1}{2(1+A^2)}\left(\frac{\tau-mT_{\mathrm{rev}}}{T_{\mathrm{collapse}}}\right)^2\right)
\label{equltimativeformel}
\end{equation}
with $\tau=kt$. In the following we want to discuss the structure of this central result. 
The most important features are the plasma oscillations ($\tilde \omega_p$), 
their collapse ($T_{\mathrm{collapse}}$) and their revivals ($T_{\mathrm{rev}}$). 

The phase $\tilde \varphi=\tilde \varphi (\tau,m)$ can be ignored for a qualitative
discussion -- it is a complicated expression and can be 
found in the appendix. Importantly, ${\tilde \varphi}$ varies slowly with time and thus needs 
only be taken into account when quantitative agreement with exact calculations 
over extremely long time scales is sought.
 
The parameter $A=A(\tau,m)=\tau\cdot \Sigma_\tau -m \cdot \Sigma_m$ 
(expressions for the constants  $\Sigma_\tau$ and $\Sigma_m$ can be found in the appendix) 
describes an additional slow broadening and decay of the signal. As for the phase
${\tilde\varphi}$, the inclusion of $A$
leads to quantitative agreement with exact calculations as shown later,  
but need not be discussed further here. 
\\
Thus we concentrate on the important plasma oscillations ($\tilde \omega_p$), 
their collapse ($T_{\mathrm{collapse}}$) and their revivals ($T_{\mathrm{rev}}$). 

The analytical formula for the generalized plasma frequency 
for arbitrary $\Lambda$ is
\begin{equation}
  \tilde{\omega}_p=\omega_p(1-2c_1g-5c_2g^2) \ , 
\label{eq_plasma_analyt}
\end{equation}
which is valid both in the Rabi and the Josephson regime. 
Here, $c_1=({1+\Lambda/4})/({1+\Lambda})$ and $c_2=({1+\frac{\Lambda}{5}+\frac{\Lambda^2}{4^2}})/(1+\Lambda)^2$ 
are $\Lambda$-dependent numbers of the order of one
and $g=\frac{UV(j_0)}{\omega_p^2}$ is a dimensionless interaction parameter. 
We give a more elaborate discussion of this expression in section \ref{sec_dis}.

For the revival time we find
\begin{equation}
  T_{\mathrm{rev}}=\frac{\pi}{U}\frac{(1+2c_1g)}{(c_1+5c_2g)} ,
  \label{eq_revival_analyt}
\end{equation}
and for the collapse time
\begin{equation}
  T_{\mathrm{collapse}}=\frac{1}{2 g\, \Delta V_0\, \omega_p(c_1+5c_2g)} \ ,
  \label{eq_collapse_analyt}
\end{equation}
with 
\begin{equation}
  \Delta V_0=\sigma V'(j_0)/V(j_0)
\end{equation}
being the width of the wavepacket in energy in units of the mean excited energy. 
Again, a more elaborate discussion of these results will be done in section \ref{sec_dis}.

\subsection{Simple Rabi limit}
In the well studied Rabi limit, i.e. when $\Lambda \ll 1$, our results simplify.
In particular, ${\tilde \omega_p} \rightarrow \omega_p$, $T_{\mathrm{rev}}\rightarrow \frac{\pi}{U}$,
and $T_{\mathrm{collapse}}\rightarrow (2 g \Delta V_0 \omega_p)^{-1}$. Moreover, only the
main off-diagonal contribution $k=1$ of the matrix $A_{nn\pm k}$ needs to be taken into account.
Thus, in the Rabi limit, the dynamics of the population imbalance is governed by
the simple expression
\begin{equation}
  j(t)=j_0 \sum_m \cos\left(\omega_p t\right) \exp \left(-2 \omega_p^2g^2(\Delta V_0)^2\left(t-\frac{\pi 
m}{U}\right)^2 \right) \ ,
\label{eq_rabilimitformel}
\end{equation}
a result that with an appropriate identification of the parameters 
can also be found in the literature \cite{PitStr01}.

\section{Comparison of results}
\begin{figure}[h!]
\includegraphics[width=80mm]{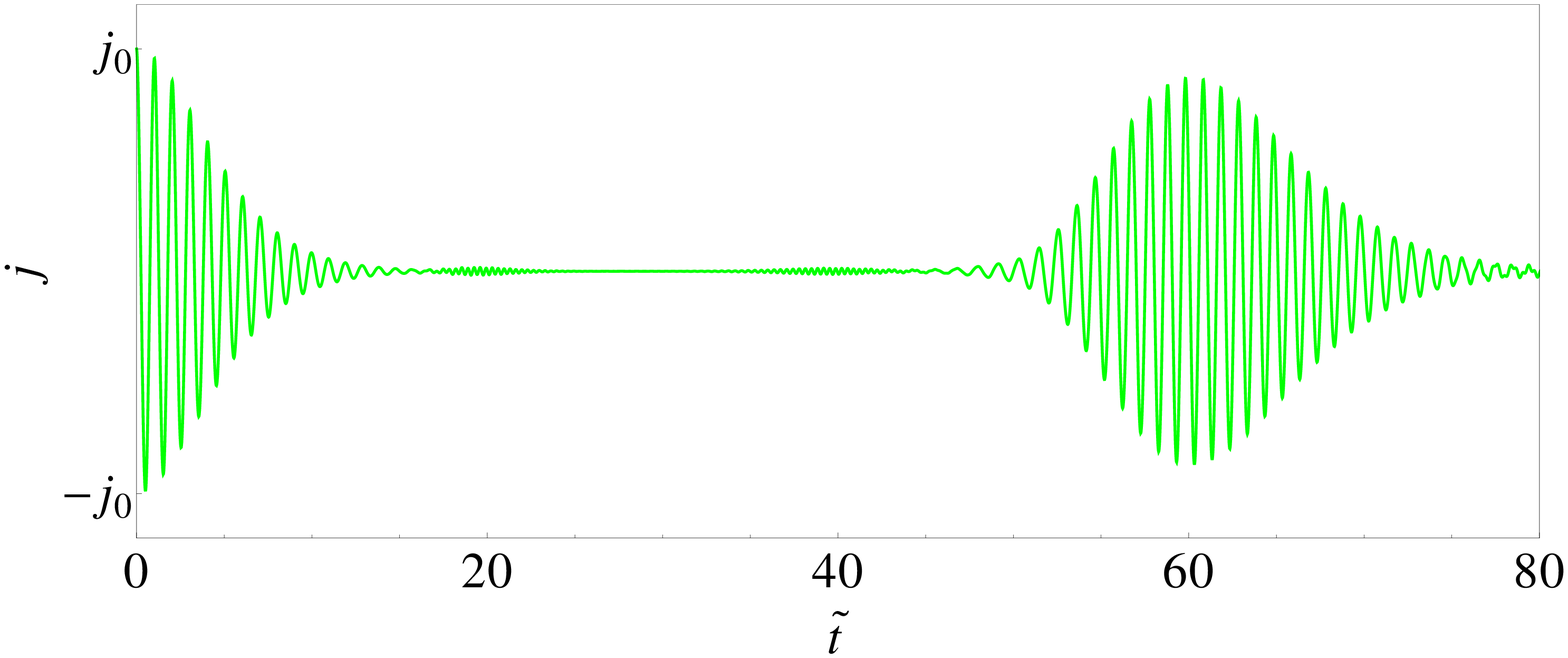} \\
\includegraphics[width=80mm]{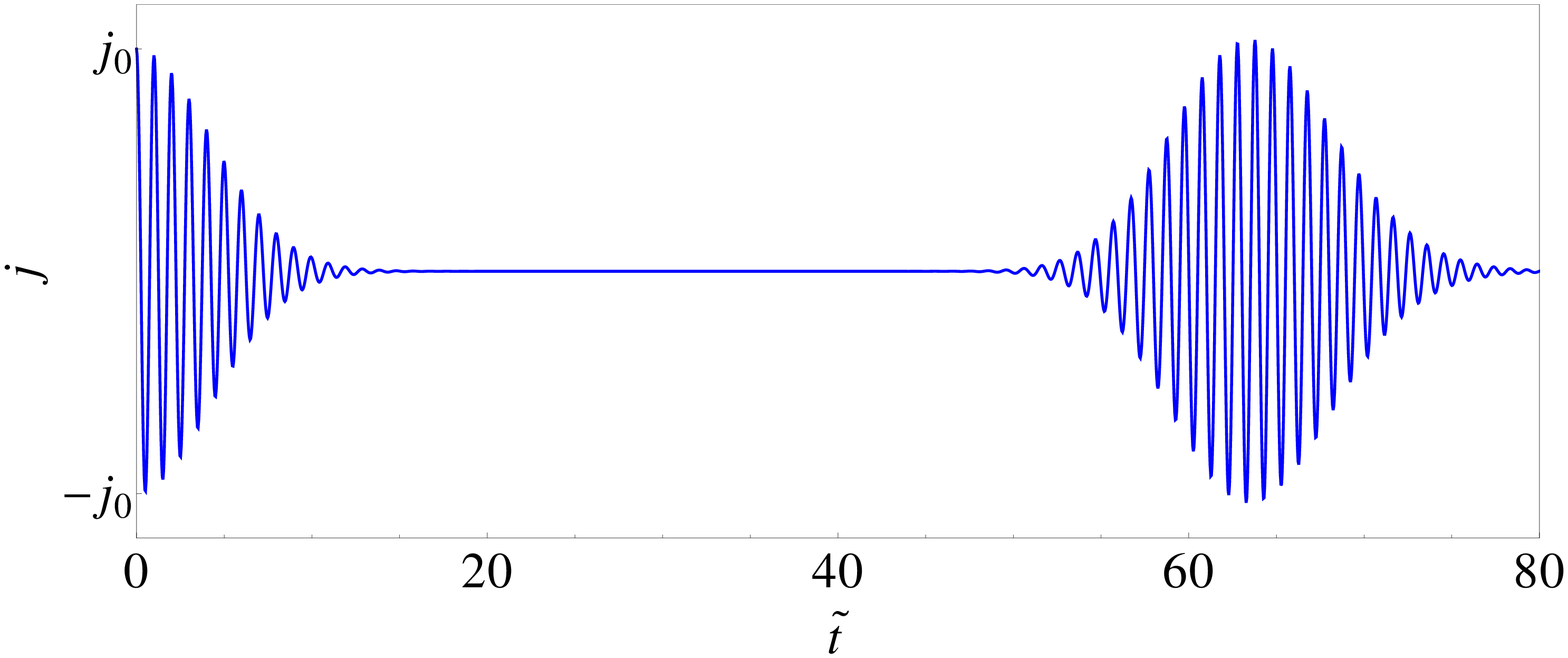}
\caption{Comparison of the exact dynamics (top) of the population imbalance $j$ with the 
improved semiclassical expression (\ref{equltimativeformel}) (bottom)
as a function of the dimensionless time $\tilde t=\tilde \omega_p t/2 \pi$ for $\Lambda=25$, $T=10$, $N=100$ 
and an initial $j_0=10$. }
\label{figdynj25}
\end{figure}

Equation (\ref{equltimativeformel}) describes the dynamics of the 
population imbalance without any free parameter.
The population imbalance oscillates with the generalized plasma frequency $\tilde \omega_p$, 
with roughly a Gaussian envelope of width $T_{\mathrm{collapse}}$ (note that the
parameter $A$ contributes to the envelope, in particular for long times). The sum over $m$ 
counts the revivals -- the initial collapse dynamics is captured by $m=0$, the first revival 
corresponds to the contribution of $m=1$, and so on. 
The sum over $k$ 
takes into account further off-diagonal contributions in the matrix $A_{nn\pm k}$ which
lead to small revivals 
 (of the order of $d_k$) at earlier times $mT_{\mathrm{rev}}/k$ with $k-$fold frequency.
 For $\Lambda = 25$, for instance, one can see tiny contributions of $k=3$ at one and two
 thirds of the full revival time in Fig. \ref{figdynj25}.
 
\begin{figure}[h!]
\includegraphics[width=80mm]{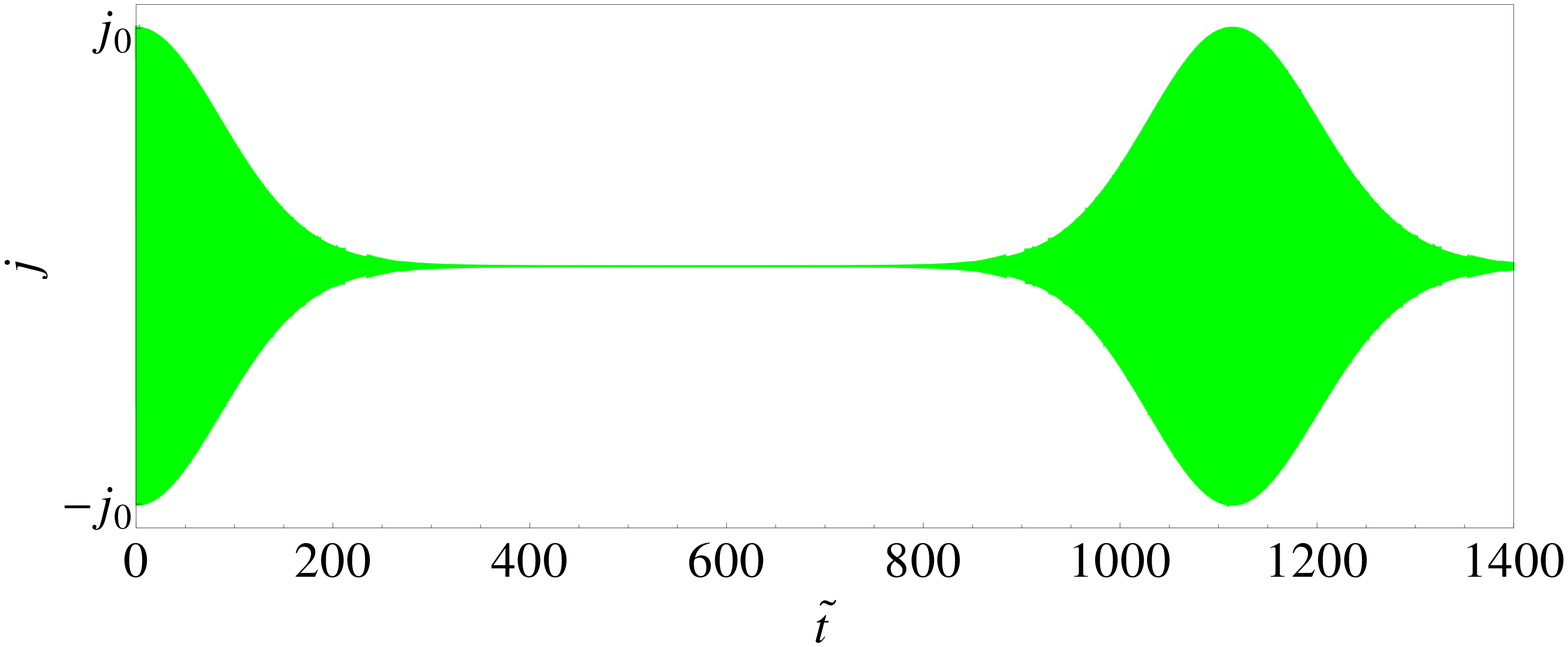} \\
\includegraphics[width=80mm]{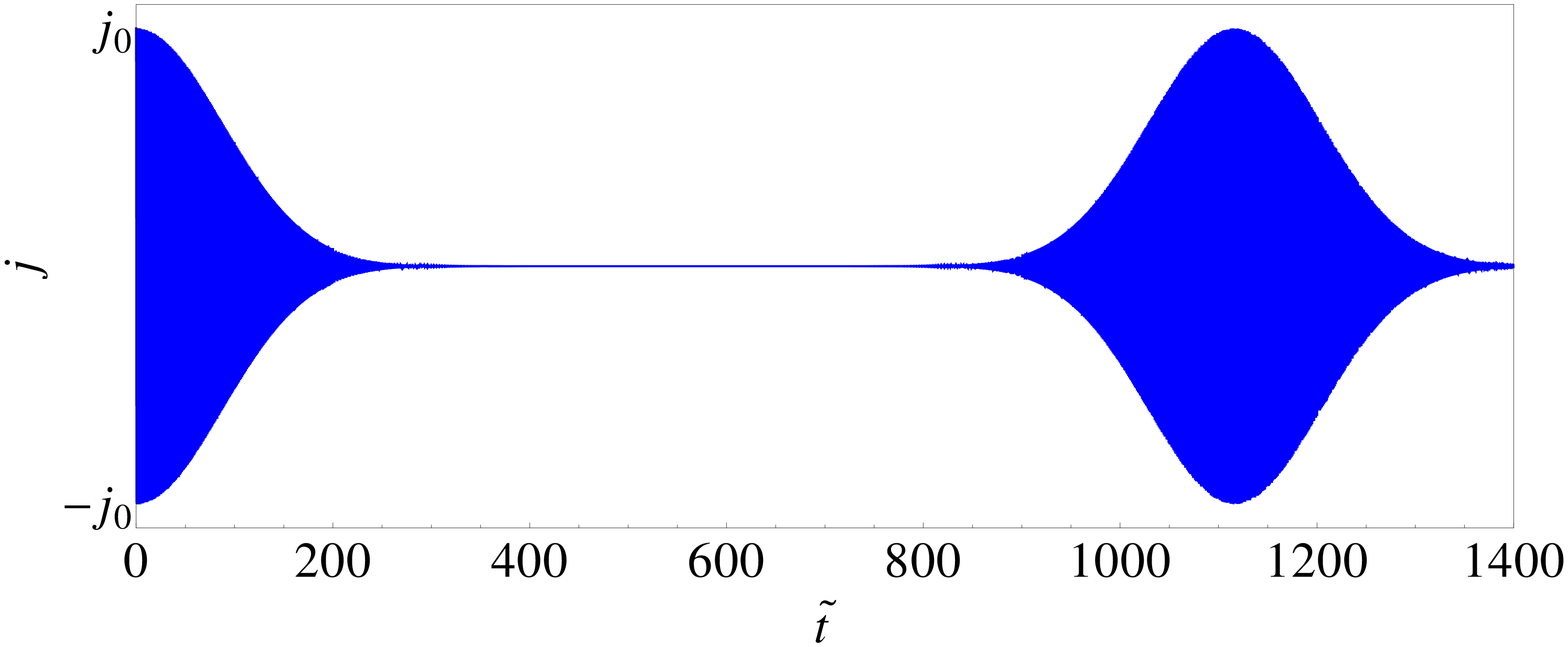} \\
\includegraphics[width=80mm]{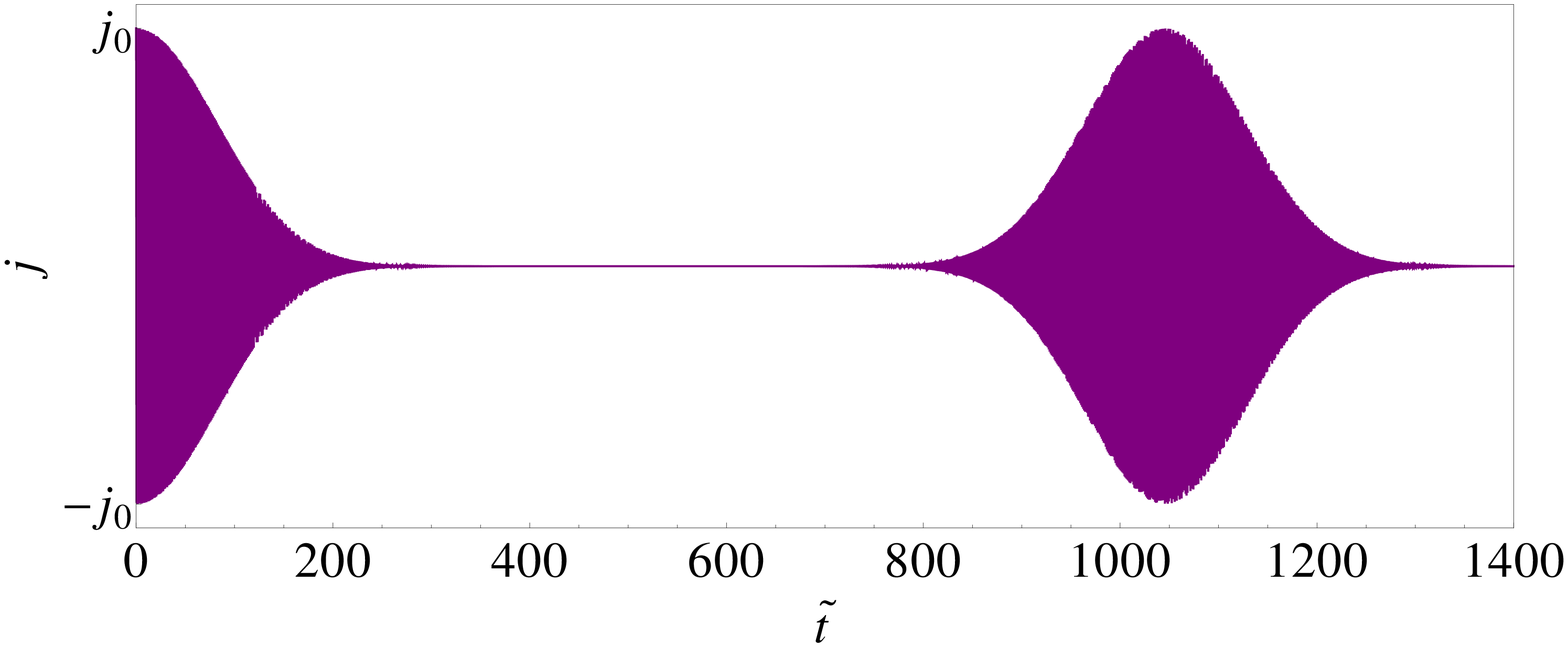}
\caption{Comparison of the exact dynamics (top) of the population imbalance $j$ with the 
improved semiclassical expression (\ref{equltimativeformel}) (middle) and the expression for the Rabi limit (\ref{eq_rabilimitformel}) (bottom)
as a function of the dimensionless time $\tilde t= \tilde \omega_p t/2 \pi$ for $\Lambda=0.1$, $T=10$, $N=100$ 
and an initial $j_0=20$. }
\label{figdynj01}
\end{figure}

\begin{figure}[h!]
\includegraphics[width=80mm]{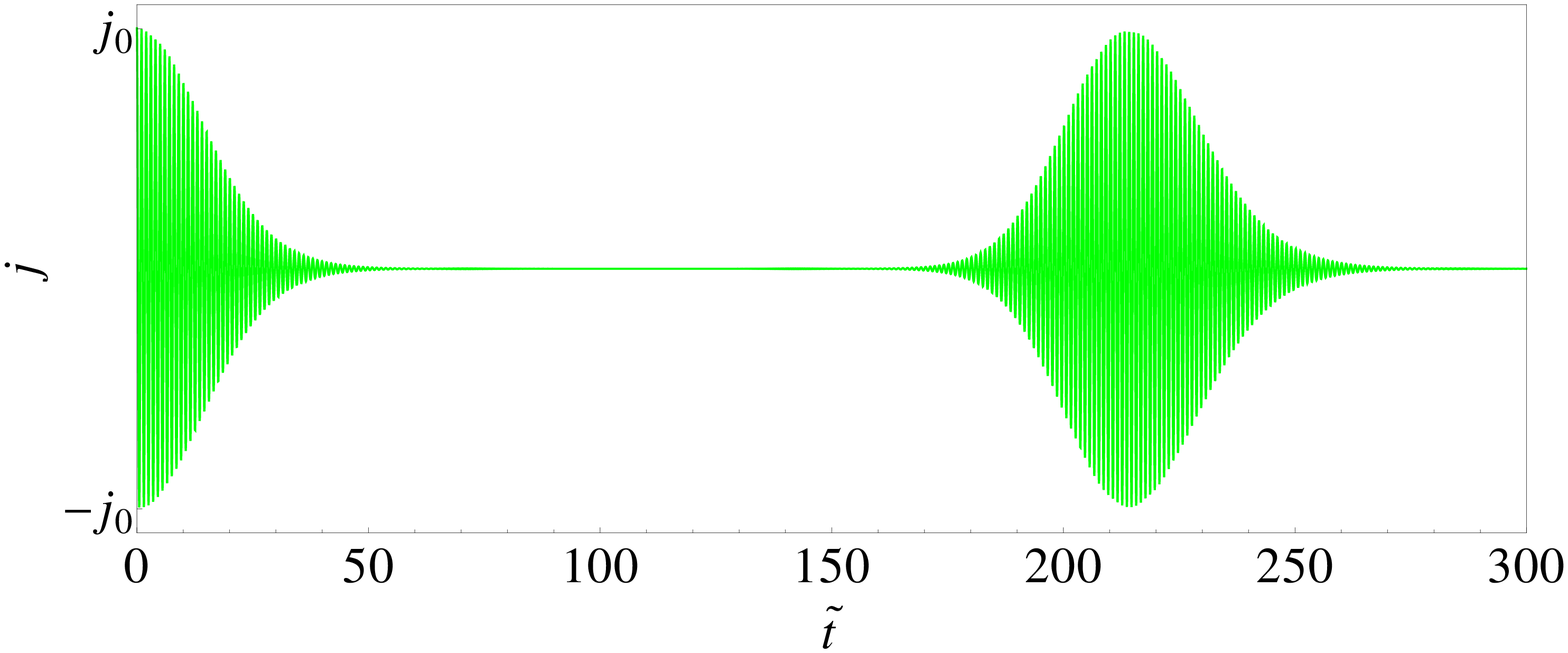} \\
\includegraphics[width=80mm]{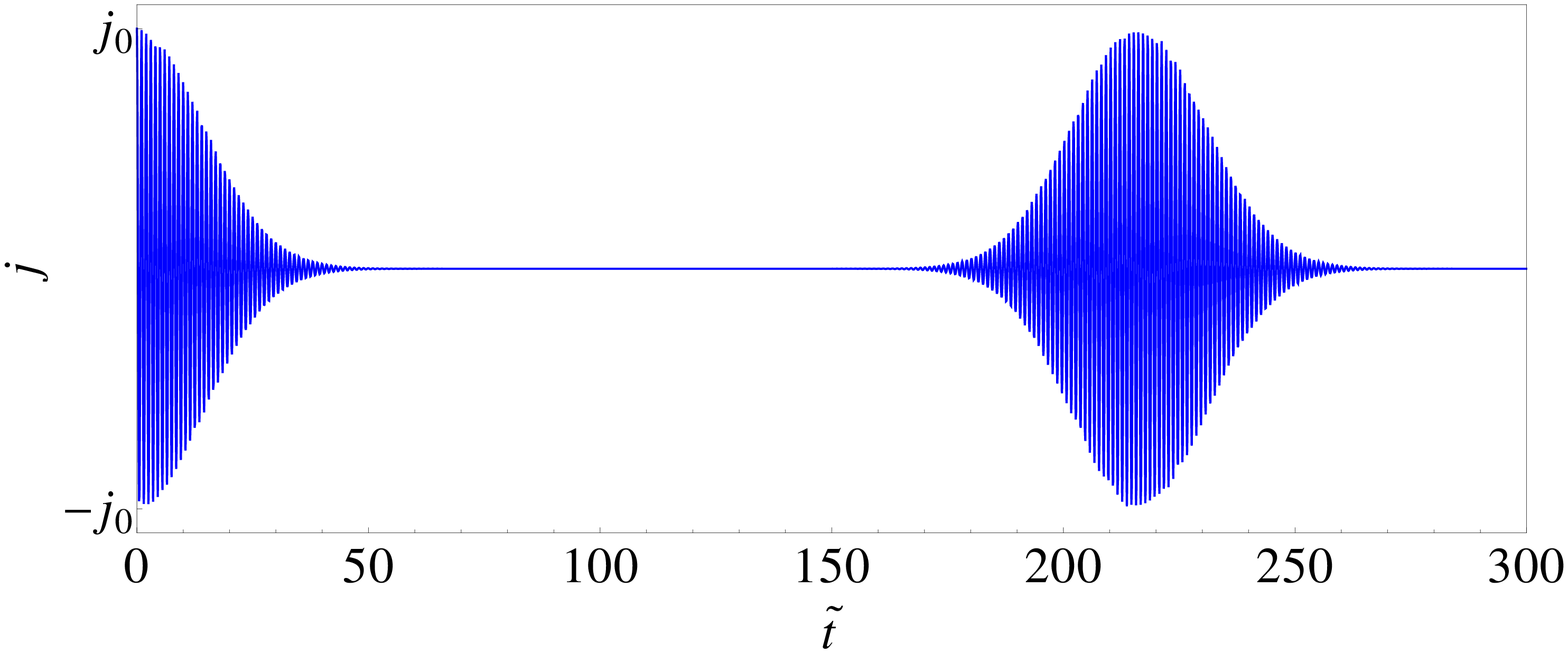} \\
\includegraphics[width=80mm]{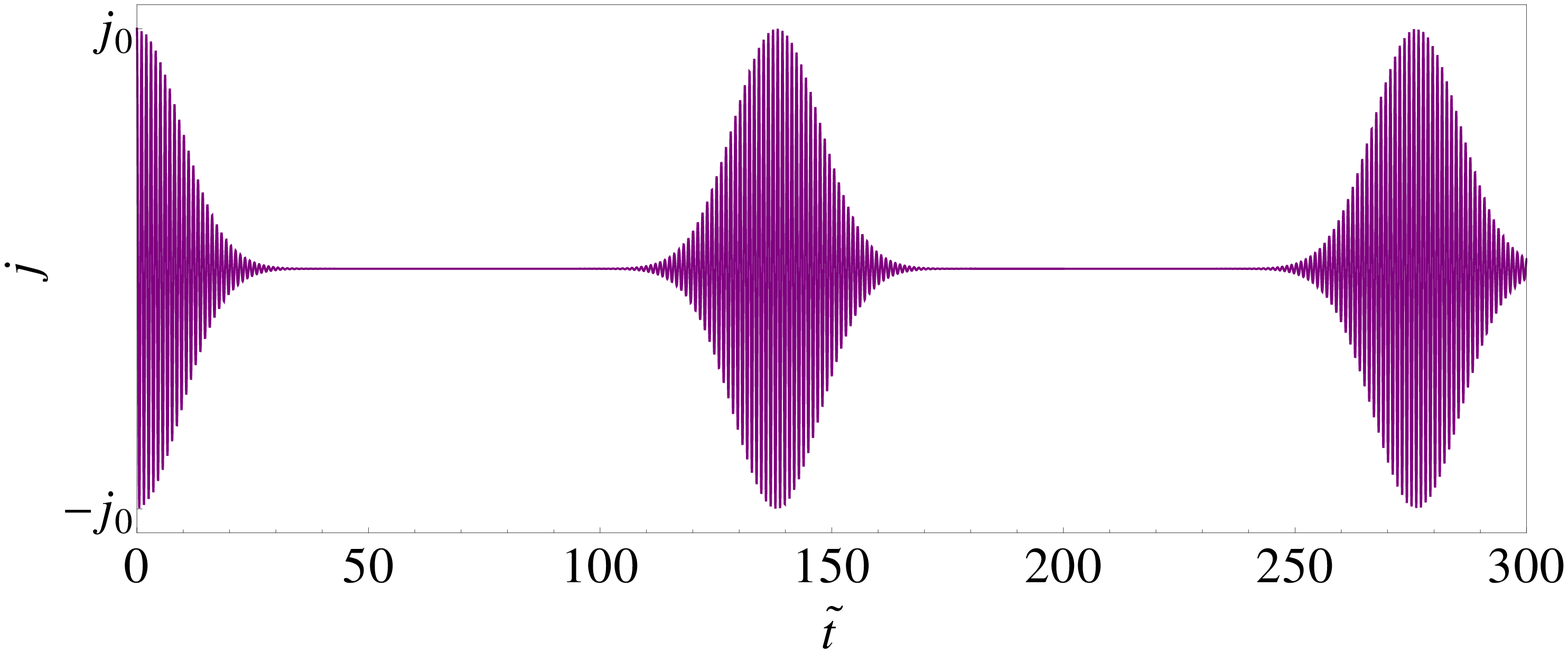}
\caption{Comparison of the exact dynamics (top) of the population imbalance $j$ with the 
improved semiclassical expression (\ref{equltimativeformel}) (middle) and the expression for the Rabi limit (\ref{eq_rabilimitformel}) (bottom)
as a function of the dimensionless time $\tilde t= \tilde \omega_p t /2 \pi$ for $\Lambda=1$, $T=10$, $N=100$ 
and an initial $j_0=20$. }
\label{figdynj1}
\end{figure}

\begin{figure}[h!]
\includegraphics[width=80mm]{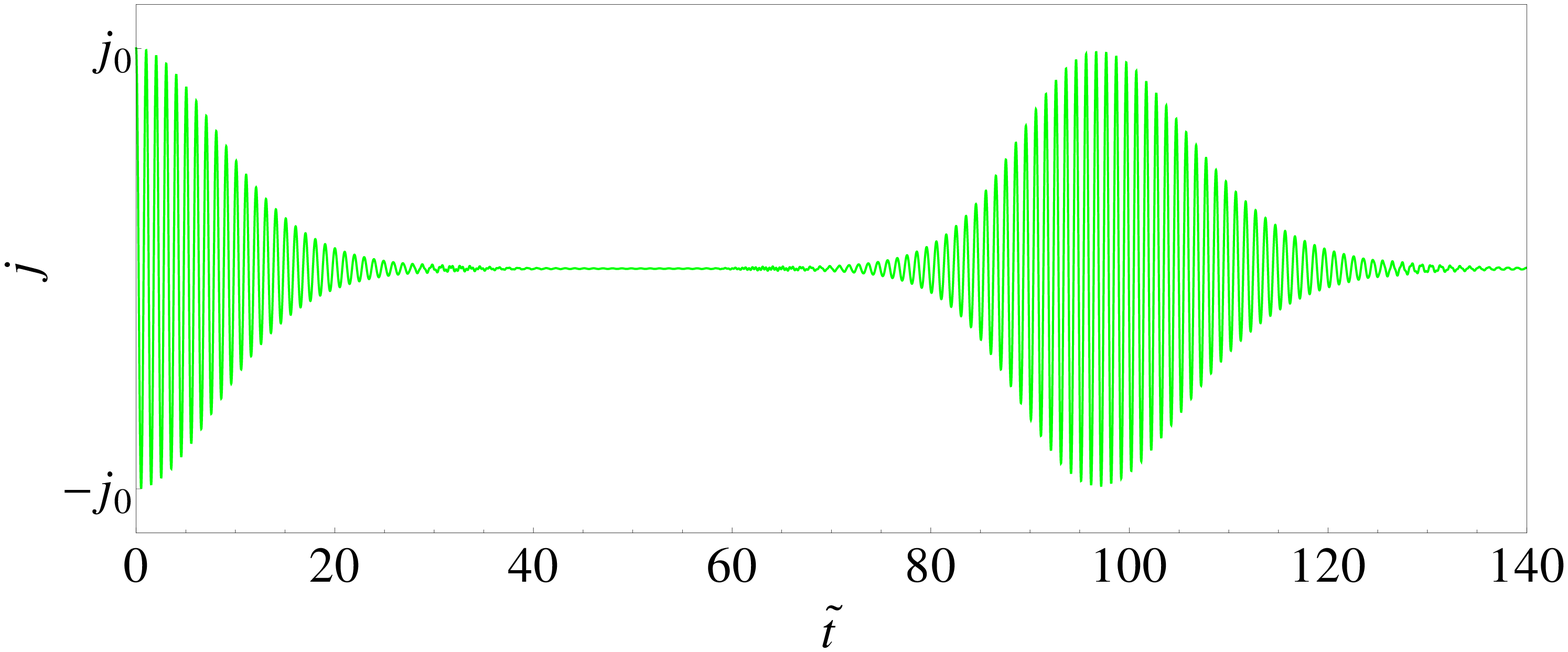} \\
\includegraphics[width=80mm]{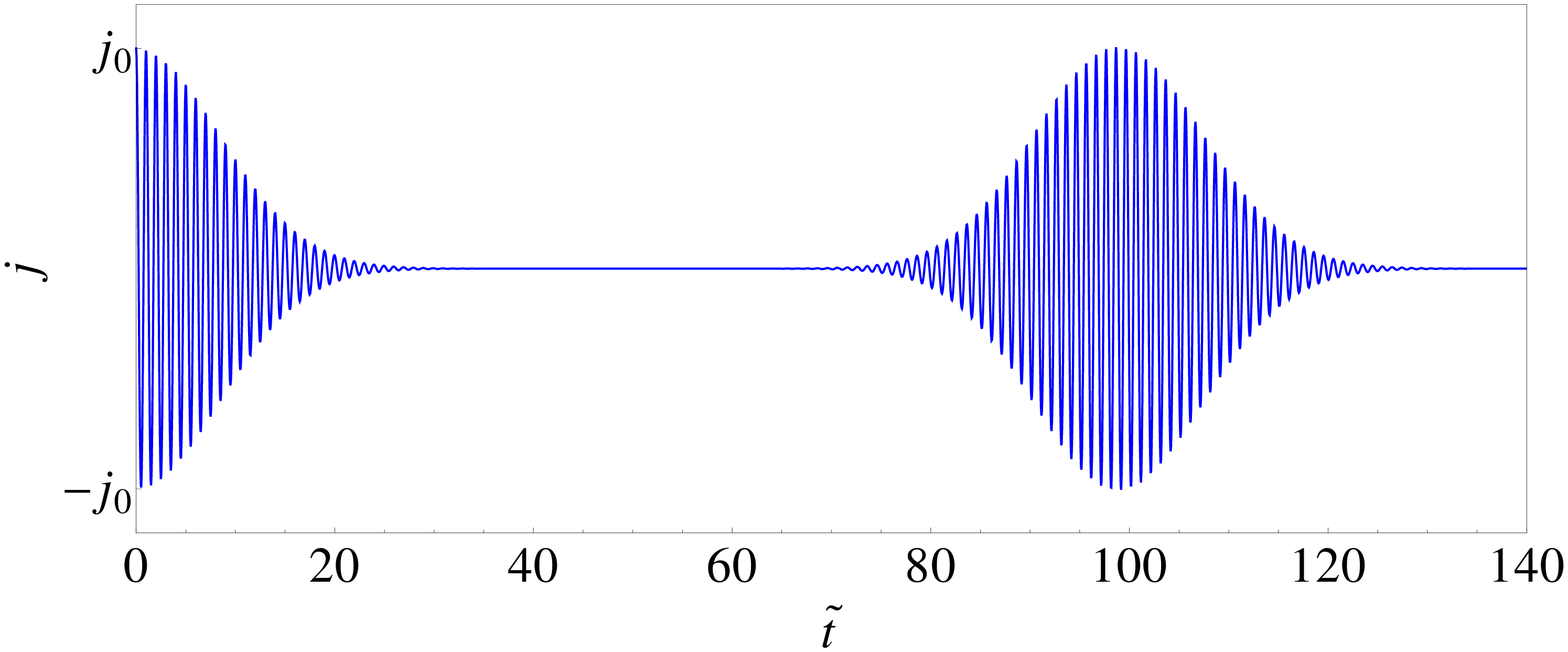} \\
\includegraphics[width=80mm]{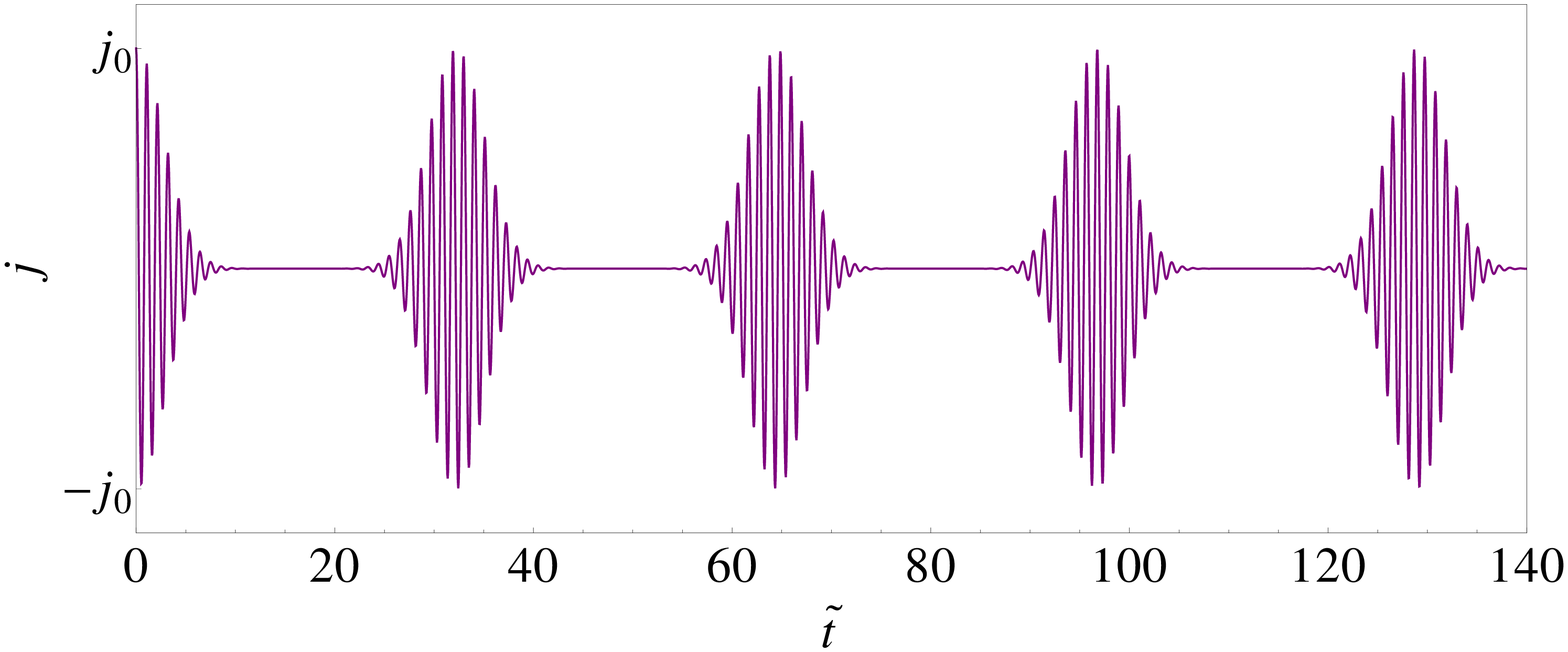}\\
\caption{Comparison of the exact dynamics (top) of the population imbalance $j$ with the 
improved semiclassical expression (\ref{equltimativeformel}) (middle) and the expression for the Rabi limit (\ref{eq_rabilimitformel}) (bottom) 
as a function of the dimensionless time $\tilde t=\tilde \omega_p t/2 \pi$ for $\Lambda=10$, $T=10$, $N=100$ 
and an initial $j_0=10$. }
\label{figdynj10}
\end{figure}

Figures \ref{figdynj01}, \ref{figdynj1}, \ref{figdynj10} and \ref{figdynj25} show
a comparison of the exact dynamics of the 
population imbalance, our analytical expression (\ref{equltimativeformel}) 
(taking into account $k=1$ only) and 
the simple expression for the Rabi limit (\ref{eq_rabilimitformel}), for different 
values of $\Lambda$ between $0.1$ and $25$.

\begin{figure}[h!]
\includegraphics[width=60mm]{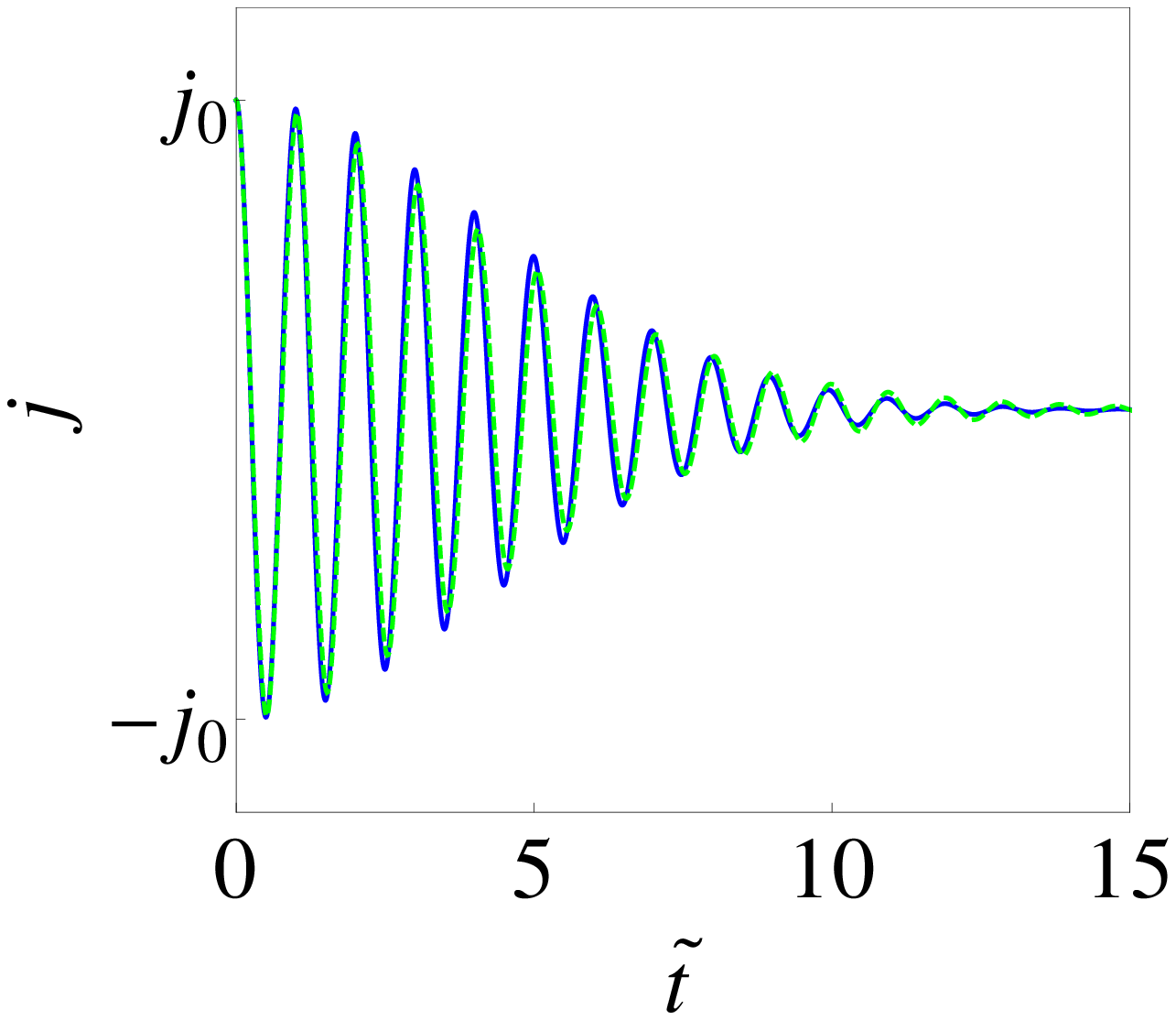}
\caption{Detailed comparison of the numerical exact initial collapse dynamics of $j$ (green, dashed line) 
and the improved semiclassical analytical results (blue, solid line) for $\Lambda=25$, $N=100$, $T=10$ and $j_0=10$ as a function of the dimensionless time $\tilde t=\tilde \omega_p t/2 \pi$. }
\label{fig_collapse_25}
\end{figure}

Obviously, our 
semiclassical expression
(\ref{equltimativeformel}) describes the exact dynamics almost perfectly 
over this huge range of values of $\Lambda$. By contrast, the simple expression (\ref{eq_rabilimitformel})
is valid, indeed, for only very small values of $\Lambda$ ($\Lambda=0.1$), as expected.
For increasing $\Lambda$ the simple Rabi expression fails, as can be seen from
Figs. \ref{figdynj1} and \ref{figdynj10}. 

\section{Discussion}\label{sec_dis}

Having an analytical expression for the time evolution of the population 
imbalance allows us to discuss the dependence of the collapse- and revival 
time and the plasma oscillation frequency on
the relevant parameters of the system.

\subsection{Plasma oscillation frequency}
The plasma oscillation frequency was found to be
\begin{equation}
\tilde{\omega}_p=\omega_p(1-2c_1g-5c_2g^2) \ 
\end{equation}
 with $c_1=({1+\Lambda/4})/({1+\Lambda})$, $c_2=({1+\frac{\Lambda}{5}+\frac{\Lambda^2}{4^2}})/(1+\Lambda)^2$,
and $g=\frac{UV(j_0)}{\omega_p^2}$.
For very small $\Lambda$, the correct ${\tilde \omega_p}$ approaches the standard plasma frequency $\omega_p$, 
since the constants $c_1$ and $c_2$ tend to one in this limit, and 
the parameter $g$ approaches zero: with $V(j_0)\approx \omega_p^2 j_0^2/(2NT)$ (harmonic approximation of the
potential), it is worth 
writing the latter parameter in the form
\begin{equation} 
 g\approx\frac{\Lambda}{8}(2j_0/N)^2
\label{eq_g_anders}
\end{equation}
which shows that $g$ tends to zero linearly in $\Lambda$ for fixed initial imbalance $(j_0/N)$. 
However, for increasing $\Lambda$ the correction terms in $\tilde \omega_p$ become
more and more relevant, especially for large $j_0$, as can be seen from (\ref{eq_g_anders}). 
\begin{figure}[htb]
\includegraphics[width=80mm]{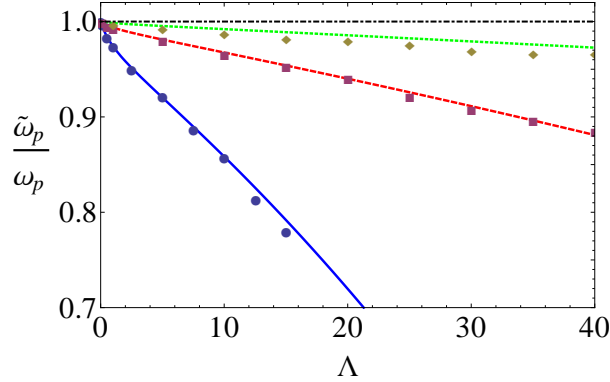}
\caption{Comparison of the numerically exact (symbols) and improved semiclassical analytical (\ref{eq_plasma_analyt})
 (lines) plasma oscillation frequency as a function of $\Lambda$ for different $j_0$: $j_0=5$: 
diamonds, dotted line, $j_0=10$: squares, dashed line, $j_0=20$: circles, solid line, 
for a total number of $N=100$ particles.
The constant dash-dotted line indicates the simple plasma frequency $\omega_p$, valid only in the
Rabi regime $\Lambda\ll 1$. }
\label{fig_omega}
\end{figure}
Figure \ref{fig_omega} shows a comparison of the plasma frequency obtained from
numerically exact results and 
the semiclassical expression (\ref{eq_plasma_analyt}) as a function of $\Lambda$ for
different initial imbalance $j_0$. 
It can be seen that the classical plasma frequency $\omega_p$ is only a good 
approximation for very small $\Lambda$, as expected. Especially for relatively
large initial imbalance
$j_0=20$ with $N=100$, the numerically exact plasma frequencies (circles) differ strongly from $\omega_p$, 
but are in very good agreement with the new semiclassical expression ${\tilde\omega_p}$ (blue, solid line). 
 Since $\Lambda=25$ and an initial $j_0/N\approx 0.15$ are typical experimental 
values \cite{Gat07}, this discrepancy becomes by all means relevant. Sure enough,
with the parameters given in \cite{Gat07}, our formula leads to $2\pi/{\tilde\omega_p}=39$ms -- which
is the experimentally observed value. By contrast, without our corrections one would find $2\pi/{\omega_p}=30$ms.

\subsection{Collapse time}
According to (\ref{equltimativeformel}), the collapse time is given by
\begin{equation}
 T_{\mathrm{collapse}}=\frac{1}{2 g \Delta V_0 \omega_p(c_1+5c_2g)}.
\end{equation}
The expression in front of the brackets (which can be identified with the 
collapse time in the Rabi regime, i.e. for small $\Lambda$) can be approximated as $N/(\Lambda \omega_p (2j_0/N))\sigma$.
Thus, assuming $\Lambda$ and $(j_0/N)$ are kept fixed so that $\sigma$ is proportional to $\sqrt{N}$,
the collapse time is proportional to $\sqrt{N}$. 
This $\sqrt{N}$-behavior has been stated before in \cite{Paw11,Par09}.
Our semiclassical formula shows, however, that this statement is only correct for the special 
case of fixed $\Lambda$ and $j_0/N$, or in the Rabi limit ($\Lambda\ll 1$). In all the
other cases, the collapse time 
depends in a nontrivial way on $N$ through $\Lambda$ and $j_0/N$. 
Figures \ref{figdynj01}, \ref{figdynj1}, \ref{figdynj10}, \ref{figdynj25}, and in detail Fig.
\ref{fig_collapse_25} show that our semiclassical expression for the collapse time is remarkably
reliable.

\subsection{Revival time}
\begin{figure}[htb]
\includegraphics[width=80mm]{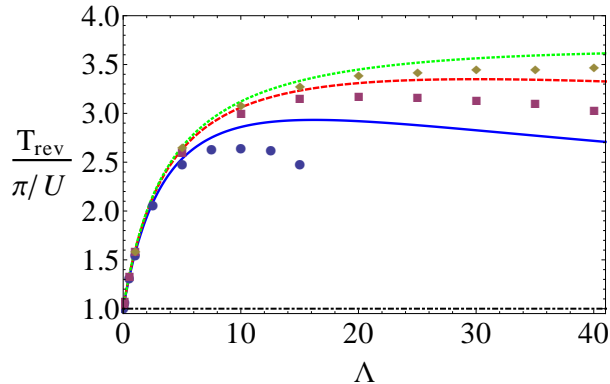}
\caption{Comparison of the numerically exact (symbols) and improved
semiclassical analytical
 revival time (lines) as a function of $\Lambda$ for different initial
 imbalance $j_0$: $j_0=5$: 
diamonds, dotted line, $j_0=10$: squares, dashed line, $j_0=20$: circles, solid line,
for a total number of $N=100$ particles. The
dashed-dotted line indicates the result $\pi/U$ of the Rabi limit, which is
obviously only valid for very small $\Lambda$. }
\label{fig_revival}
\end{figure}

Following (\ref{equltimativeformel}), the revival time is
\begin{equation}
 T_{\mathrm{rev}}=\frac{\pi}{U}\frac{(1+2c_1g)}{(c_1+5c_2g)} \ .
\label{eqrevtime}
\end{equation}
Most interestingly, in the Rabi limit it becomes independent of the number of 
particles and in fact independent of any other system parameter except
the interaction strength $U$. 
The revival time was already discussed in \cite{Ton05} where it
was found to be equal to $4 \pi$, with an 
interaction strength of $\frac{1}{4}$ (considering the different 
definition of parameters), which we confirm here, in the Rabi limit.
Furthermore, it is stated in \cite{Ton05, Paw11} that the revival time grows 
linearly with the number of particles $N$. This is obviously true
for those investigations with $UN=$ const only, as can be seen
from our expression (\ref{eqrevtime}). However, note that
even only slightly away from the Rabi limit, when $\Lambda$ approaches or becomes
greater than one, the constants $c_1$ and $c_2$ and the parameter $g$ 
become relevant. This can be seen from Fig. \ref{fig_revival}. Thus,
for $\Lambda>1$ no simple scaling law for the revival time exists.

The figure shows that for increasing $\Lambda$ the exact revival times 
differ strongly from the revival time $\pi/U$ predicted by the Rabi limit formula. 
On the other hand, it can be seen that the improved semiclassical expression 
(\ref{eqrevtime}) reproduces the exact revival times very nicely even for
$\Lambda>1$. For increasing values of $\Lambda$ 
 the self-trapping fraction of the phase space is increasing as well, 
such that for large initial excitations, e.g. $j_0=20$ for $N=100$,
the semiclassical analysis ceases to give reliable results.

\section{Conclusion}
We applied semiclassical methods to the well-known two-mode Bose-Hubbard model, in 
order to investigate in detail BEC tunneling in a 
double-well trap. Within the plasma oscillation regime
we found analytical expressions for the energy spectrum and the initial 
state agreeing nicely with numerically exact results. 
Employing the reflection principle and the Poisson summation 
formula led us to an analytical expression for the time evolution of the population 
imbalance of the Bose gas in the double well.
This allows us to discuss the dependence of characteristic quantities of the 
dynamics, like plasma oscillation frequency, collapse and revival times, 
on the relevant system parameters. Remarkably enough, despite a wealth of
publications on the two-mode model, such detailed understanding has not been
achieved before.
Finally, our generalized formula for the plasma oscillation frequency agrees
perfectly well with experimental findings. Challenging as it may be, we hope
that our predictions for collapse and revival times will be confirmed experimentally,
too.

Semiclassical methods are well suited to study the non-equilibrium dynamics of
a Bosonic interacting many-body quantum system. For systems with more degrees of
freedom, an explicitly time dependent approach might prove useful.

\section*{Acknowledgments}
We thank Markus Oberthaler for a nice discussion.
L. S. acknowledges support from the International Max Planck Research
School (IMPRS), Dresden.

\begin{appendix}
\section{Parameters}
In order to complete the discussion of our semiclassical analytical result for the time evolution of 
he population imbalance (\ref{equltimativeformel}), we present here the definition of the
remaining parameters. The phase of the oscillation reads
\begin{equation}
 \tilde \varphi = - m 
\varphi_m+\frac{1}{2}\arctan A- \frac{A}{2(1+A^2)}\left(\frac{\tau-mT_{\mathrm{rev}}}{T_{\mathrm{collapse}}}\right)^2 \ ,
\end{equation}
where the dominantly $m$-dependent part is defined separately as
\begin{equation}
  \varphi_m=2\pi\bar{V}(1+c_1g-1/(2\bar V)) \ .
\end{equation}
$\bar V$ is the mean excited energy in units of the plasma frequency
\begin{equation}
  \bar V =V(j_0)/\omega_p \ .
\end{equation}
Furthermore, the quantity
\begin{equation}
  A=\tau \Sigma_\tau -m \Sigma_m \ 
\end{equation}
contributes to an overall slow spread and decay of the signal. It can be
separated in a $\tau$- and a $m$ dependent contribution with
\begin{equation}
  \Sigma_\tau=10c_2(\Delta V_0)^2 g^2 \omega_p
\end{equation}
and
\begin{equation}
  \Sigma_m=4 \pi c_1 (\Delta V_0)^2 g \bar V \ .
\end{equation}
For completeness, we repeat the expressions for
$c_1=({1+\Lambda/4})/({1+\Lambda})$, $c_2=({1+\frac{\Lambda}{5}+\frac{\Lambda^2}{4^2}})/(1+\Lambda)^2$,
$g=\frac{UV(j_0)}{\omega_p^2}$, and $\Delta V_0=\sigma V'(j_0)/V(j_0)$ from section \ref{subsecpop}.

\end{appendix}


\begin{thebibliography}{99}
\bibitem{BraunFootnote} Note that in Braun \cite{Bra93} $\phi$ is introduced as a momentum
operator.
\bibitem{NFootnote} Strictly speaking, 
the argument of the square root should read $(N+1)^2/4-j^2$, which in 
the semiclassical limit may well be replaced by our expression.
\end{thebibliography}
\end{document}